\documentclass{aa}  
\usepackage{booktabs}
\usepackage{amsmath}
\usepackage{amssymb}
\begin{document}

\title{The peculiar nebula Simeis 57}

\subtitle{II. Distance, nature and excitation}

\author{L.H.T. Oudshoorn\inst{1}
      \and
      F.P. Israel\inst{1}
      \and
      J. Brinchmann\inst{1,2}
      \and
      M.B.C. Kloppenburg\inst{1,3}
      \and
      A.G.A. Brown\inst{1}
      \and
      J. Bally\inst{4}
      \and
      T.R. Gull\inst{5}
      \and 
      P.T. Boyd\inst{5}
      }
\let\oldAA\AA
\renewcommand{\AA}{\text{\normalfont\oldAA}}
\institute{Sterrewacht Leiden, P.O. Box 9513, 2300 RA Leiden, The Netherlands
     \and
         Instituto de Astrof\'isica e Ci\^encias do Espa\c{c}o, Universidade do Porto, CAUP, Rua das Estrellas, PT4150-762 Porto, Portugal
    \and
         Presently: Pels Rijcken, P.O Box 11756, 2502 AT Den Haag, The Netherlands
    \and 
        Center for Astrophysics and Space Astronomy (CASA), Univ. of Colorado, 389 UCB, Boulder, CO 808309, USA
    \and 
        NASA/GSFC, Mail Code: 667, Greenbelt, MD 20771, USA
    }

\date{Received ***; accepted 22-04-2021}

% \abstract{}{}{}{}{} 
% 5 {} token are mandatory

\abstract
% context heading (optional)
% {} leave it empty if necessary  
{ Simeis~57 (HS~191) is an optically bright nebula in the Cygnus X
  region with a peculiar appearance that suggests an outflow from a
  rotating source. Newly obtained observations and archival data
  reveal Simeis~57 as a low-density ($n_{e}\,\sim\,100$ cm$^{-3}$)
  nebula with an east-to-west excitation gradient. The extinction of
  the nebula is $A_{V}\,\leq$ 2 mag.  The nebula is recognizable but
  not prominent in mid- and far-infrared images. In its direction,
  half a dozen small CO clouds have been identified at $V_{LSR}$ = + 5
  km s$^{-1}$. One of these coincides with both the optical nebula and
  a second CO cloud at the nebular velocity $V_{LSR}\,\approx$ -10 km
  $^{-1}$. No luminous stars are embedded in these molecular clouds,
  nor are any obscured by them and no sufficiently luminous stars are
  found in the immediate vicinity of the nebula. Instead, all
  available data points to the evolved star HD~193793 = WR~140 (an
    O4-5 supergiant and WC7 Wolf-Rayet binary) as the source of
  excitation, notwithstanding its large separation of $50'$,
  about 25 pc at the stellar distance of 1.7 kpc. Simeis~57 appears
  to be a part of a larger structure surrounding the HI void centered
  on HD~193793.}


   \maketitle
%
%-------------------------------------------------------------------

\section{Introduction}
The Galactic nebula Simeis~57, also known as HS~191 (\citet{Gaze1951},
\citet{Gaze1955}) is a very bright emission nebula of peculiar shape
in the constellation of Cygnus \citep[cf.][]{Parker1979}. Often
referred to as the Propeller Nebula, this nebula is a popular object for
amateur astrophotographers. Its major features are two curved
nebulosities (DWB~111 and DWB~119, Dickel et al. 1969). 
The center of this nebula is
cut by a long dust filament that continues adjacent to the nebular
patch DWB~118. Other nearby nebulae are DWB~126 to the north and
DWB~108 and DWB~107 to the south; the latter has the appearance of a
bright rim. At a Galactic longitude of $80.3^{\circ}$, Simeis~57 is
inside the solar circle but its actual distance is unknown. In this
direction, the Galactic line of sight is tangential to the
Orion-Cygnus spiral arm and several kiloparsecs long. The
relatively high Galactic latitude of $+4.7^{\circ}$ and the large
angular extent ($\sim20'$) suggest, however, that it is not very distant.

Simeis~57 is located near the edge of a large
($18^{\circ}\times13^{\circ}$) X-ray structure known as the Cygnus
superbubble \citep{cash1980} but none of the large-scale X-ray or
radio continuum maps presented by \citet{uyaniker2001} show anything
remarkable concerning its position.
Simeis~57 is far away from the nearest OB associations 
\citep[Cyg OB2 and Cyg OB8 at heliocentric distances of 1.4 kpc and 1.9 kpc, cf.][]{Rygl2012,Mahy2015}.

In an earlier paper, \citet{israel2003} presented high-resolution
radio continuum maps from which they concluded that the radio emission
of Simeis~57 is free-free thermal emission originating in a gas of
moderate electron densities with S and N2 spectra that extend 
a modest foreground extinction. All of this points to excitation by 
a star that is sufficiently bright to be easily identified, but the 
lack of an obvious candidate meant that the nature of Simeis~57 
was left a mystery.

\begin{figure}
\includegraphics[width=0.48\textwidth]{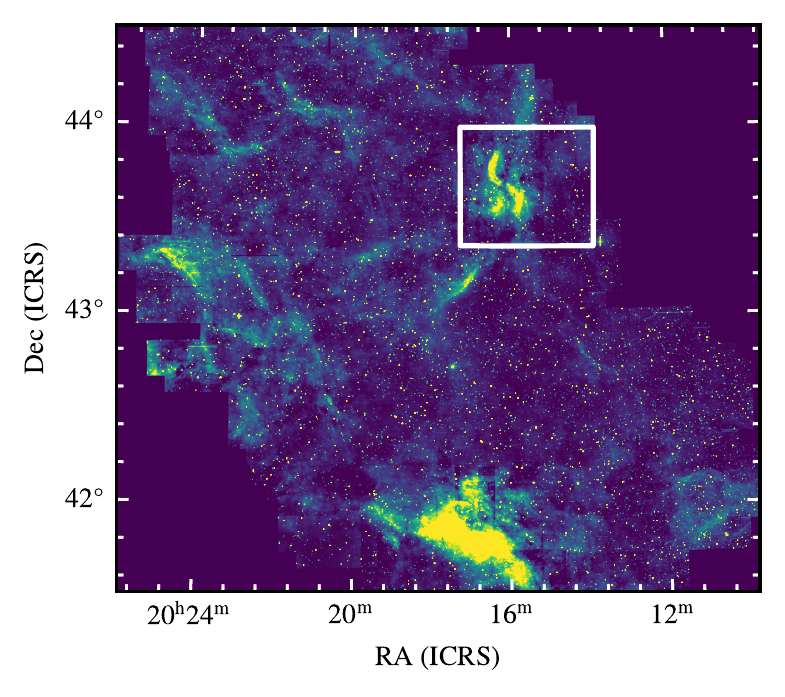}
\centering
\caption{Mosaic of H$\alpha$ line emission in the $3^{o} \times 3^{o}$
  region surrounding Simeis~57, centered on $\alpha=20^{h}17^m50.4^s$,
  $\delta=+43^{\circ}01'48"$. The white box indicates the region
  studied in this paper, which is enlarged in Fig.\ref{fig:Ha_overview}}
\label{fig:Ha_mosaic}
\end{figure}

\begin{figure}
\includegraphics[width=0.48\textwidth]{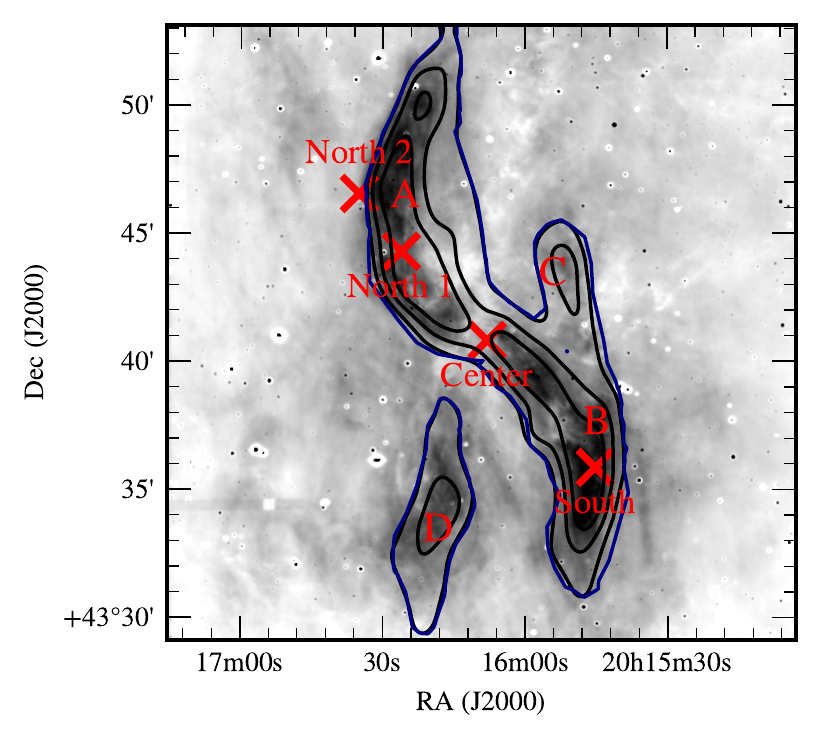}
\centering
\caption{H$\alpha$ line emission map of Simeis 57
  % divided into four
  with the subregions A, B, C, and D. The black contours correspond to
  DRAO 1420 MHz radio continuum flux densities of 20, 25, and 30
  $\mathrm{mJy \, arcmin^{-2}}$. The red crosses denote the central
  positions of the IDS long-slit spectra from Table~1 and
  Fig\,\ref{fig:spectra_3incol}.
  }
\label{fig:Ha_overview}
\end{figure}

In this work, we investigate the ionized gas, dust, and stars in the
field of Simeis~57 in a further attempt to identify its nature and the
source of its excitation. Fig.\,\ref{fig:Ha_mosaic} shows part of the
{\it Isaac Newton Telescope} Photometric H-Alpha Survey \citep[IPHAS,
][]{iphas1} combined with the new H$\alpha$ data described below.
Simeis~57 stands out in brightness and shape among the filaments
at the outskirts of the Cygnus-X region. Fig.\,\ref{fig:Ha_overview} 
shows the division of the nebula into
subregions A, B, C, and D \citep[cf.][]{israel2003} with respective
 surface areas of 45, 30, 10, and 15 $\mathrm{arcmin}^2$ within the
lowest contour. Regions A (DWB~111) and B (DWB~119) form the
S-shaped nebula, region C is a spur that extends northward from
region B, and region D (DWB~118) is a separate filament to the
southeast.
% Figure 1: Ha large field overview

%Figure 2: Ha small field 1420 MHz overlay with ABCD identification

\section{Observations}

% Figure 3: spectra

\begin{figure*}
\includegraphics[width=0.98\textwidth]{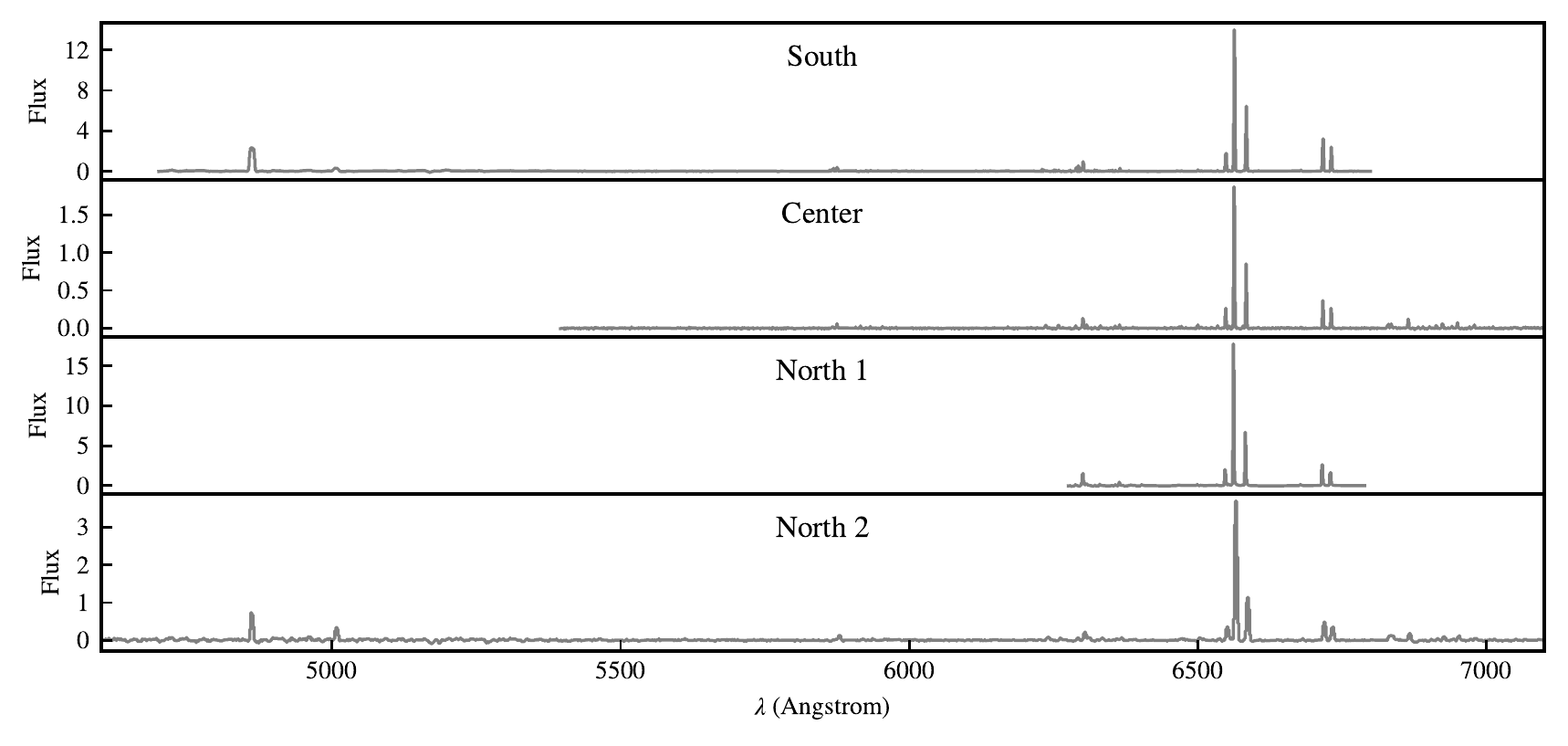}
\centering
\caption{Spatially integrated long slit spectra at the positions indicated in
  Fig.\,\ref{fig:Ha_overview}. Fluxes are in units of
  $10^{-16}\mathrm{erg\, cm^{-2} s^{-1}\AA^{-1}}$. }
\label{fig:spectra_3incol}
\end{figure*}

% Figure 4 Line images with radio contours

\begin{figure*}
\includegraphics[width=0.98\textwidth]{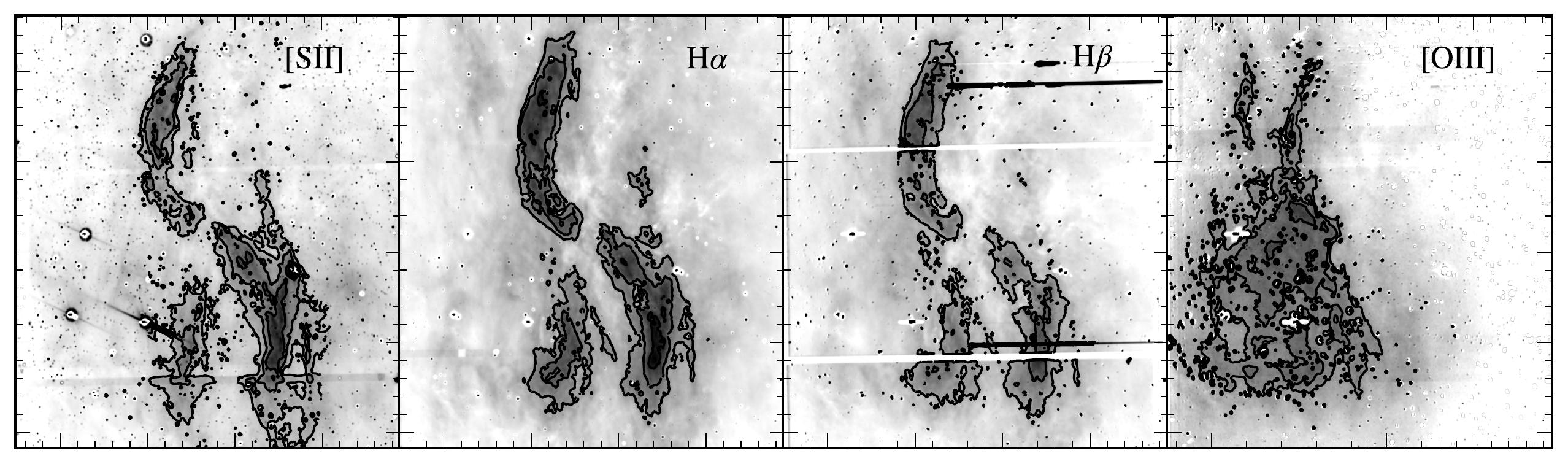}
\centering
\caption{Continuum-subtracted line images of Simeis 57 in identical
  fields of view. From left to right: [SII](6716+6730),
  H$\alpha$+[NII], H$\beta$, and [OIII]5007.  In these images, the
  first contour corresponds to
  $I_{\lambda} = (5.6, 18, 5, 1.4) \times 10^{-17}\, \mathrm{erg
    s^{-1} cm^{-2} \, arcsec^{-2}}$, respectively. Subsequent contours
  increase by factors of 1.5. The contours in the H$\alpha$ +[NII] map
  represent the pure H$\alpha$ flux. The H$\beta$ image was not
  dithered and suffers from camera artifacts and gaps between the
  CCDs.}
\label{fig:line_images_4onrow}
\end{figure*}

\subsection{Long-slit spectra}

We extracted archival spectra taken on July 22, 1990 and May 6, 2007
and obtained new spectra on May 22-24, 2017 (Fig.\,\ref{fig:spectra_3incol},
Table\,\ref{tab:observing_log_spectra}) with the Intermediate
Dispersion Spectrograph (IDS)
on the Isaac Newton Telescope (INT)   The INT is a 2.54m telescope
located at the Observatorio del Roque de los Muchachos on the island
of La Palma (Spain). The IDS is a long-slit spectrograph located at
the Cassegrain focus that has a full
slit-length of $3.3'$ and a spatial scale of $0.4"$/pix.  The
2007 and 2017 spectroscopy used the R300V, R400V, and R600R
gratings, which have spectral resolutions of $1.87$\AA/pix,
$1.41$\AA/pix, and $0.94$\AA/pix, respectively, and a wavelength
coverage correspondingly decreasing. The 1990 spectroscopy used
Grating 10, which has a spectral resolution of $1.03$\AA/pix.  The positions
are shown in Fig.\,\ref{fig:Ha_overview}. We note that the northernmost
spectrum (N2) is right at the edge of DWB~111.

% Table 1:  Observations log

\begin{table}[h]
{\small %
\caption{Observing log of the INT-IDS spectroscopy.}
\label{tab:observing_log_spectra}
\begin{tabular}{lllllr}
\toprule
  Grating  & Obs.       & RA(2000)   & DEC(2000) & Pos.& $T_{exp}$ \\
           & date       & hh:mm:ss.s &  dd:mm:ss &     & sec \\
\hline
R600R      & 2017-05-23 & 20:15:45.2 & +43:35:52 & S   & 600 \\
R400V      & 2017-05-25 & 20:15:45.2 & +43:35:52 & S   & 600 \\
R600R      & 2017-05-24 & 20:16:08.0 & +43:40:50 & C   & 600 \\
R300V      & 2007-05-06 & 20:16:35.0 & +43:46:34 & N2  & 300 \\
Gr.\,10    & 1990-07-22 & 20:16:26.1 & +43:44:18 & N1  & 4500\\
  \hline
\end{tabular}
} %
\end{table}

\par
The wavelength solutions were derived from arc lamp spectra and
cosmic rays were removed with the L.A.Cosmic package by
\cite{2012ascl.soft07005V}. For the flux calibration of the spectra
obtained in 2017 we used the standard star SP1550$+$330. The R300V
grating spectrum was calibrated with the standard star Kopff27. The
spectra were spatially integrated over the aperture, excluding the 
outer 25 pixels on either side. The two spectra taken at the southern 
position (S) were concatenated to yield a single spectrum.

\subsection{WFC Images}

In 2016 April and 2017 April we obtained several images of Simeis~57
(Fig.\,\ref{fig:line_images_4onrow}) with the Wide-Field Camera (WFC)
at the prime focus of the INT.  With four CCDs that have $2048\times 4096$ 
pixels each, the WFC is a mosaic camera
with a CCD pixel scale of $0.33"$ and a field of view of
$34.2'\times34.2'$.  Inter-chip gaps are about $1'$ in size.

\par
For the line emission we used the standard INT narrowband filters
(H$\alpha$, FWHM $95\mathrm{\AA}$; [SII](6716+6730), $80\mathrm{\AA}$;
H$\beta$ $30\mathrm{\AA}$; [OIII]5007, $100\mathrm{\AA}$) and for
the continuum from stars and sky the broadband filters ($R$, FWHM
$1347\mathrm{\AA}$, $G$, FWHM $1285\mathrm{\AA}$).
We made multiple exposures in each filter.  Total exposure times
varied from 850 s for $R$ to 6400 s for [OIII]5007. Most lines were
exposed for about 3000 s. While taking the [SII] and
[OIII] images in 2017, we applied a random dither of a few arcmin 
to fill the inter-chip gaps.  No dithering was applied to the
2016 H$\alpha$ and H$\beta$ images. 
We eliminated the inter-chip gaps in the H$\alpha$image by combining our 
own five H$\alpha$ images with multiple exposures from the IPHAS H$\alpha$
survey \citep{iphas1}.  The average seeing of our observations is
$1.7''$, with the exception of H$\beta$ ($2.3''$) and H$\alpha$
($1.3''$). 

\par The individual exposures were debiased, flat-fielded, and combined
using THELI \citep{SchirmerTheli,ErbenTheli}. The observations were
made during bright moon and suffered from significant sky
brightness. We estimated sky levels by taking the median of selected
regions outside the nebula after masking objects using SExtractor
\citep{1996A&AS..117..393B}.  We computed astrometric solutions using
SCAMP \citep{2006ASPC..351..112B} with the Gaia DR2 catalog (see
Sect. 2.4). The astrometry should be better than 0.5$"$.

\par No spectrophotometric calibration stars were recorded during the
observations. Instead, H$\alpha$ and $r$ images are directly
calibrated with the INT/WFC Photometric H$\alpha$ survey of
the Northern Milky Way \citep[IPHAS DR2,][]{iphas1} release, which was
made with the same instrument. This dataset has $i$, $r$, and H$\alpha$
magnitudes for nearly 5000 objects in our field of view. Using
SExtractor, we extracted magnitudes of all sources with S/N$\geq7$. We
used aperture photometry with the same diameter as the IPHAS database
($2.3"$, or 7 pixels). Local background subtraction was done
with a rectangular mesh of 16 pixels around the star. The magnitude 
zero-points have errors less than 0.15 mag in $r$ and H$\alpha$.
% caused mainly by small sky differences between different nights and images.

\par No calibrators nor reference catalogs are available for the
continuum-subtracted [SII], [OIII] and H$\beta$ images. Instead, we
calibrated the relative fluxes of the images with the absolute
fluxes from the spectra in Fig.\,\ref{fig:spectra_3incol}
at the four position sampled. For H$\alpha$, the two methods agree within $20\%$.

% Figure 5: Simeis 57 fields from UV to FIR

\begin{figure*}
\includegraphics[width=0.92\textwidth]{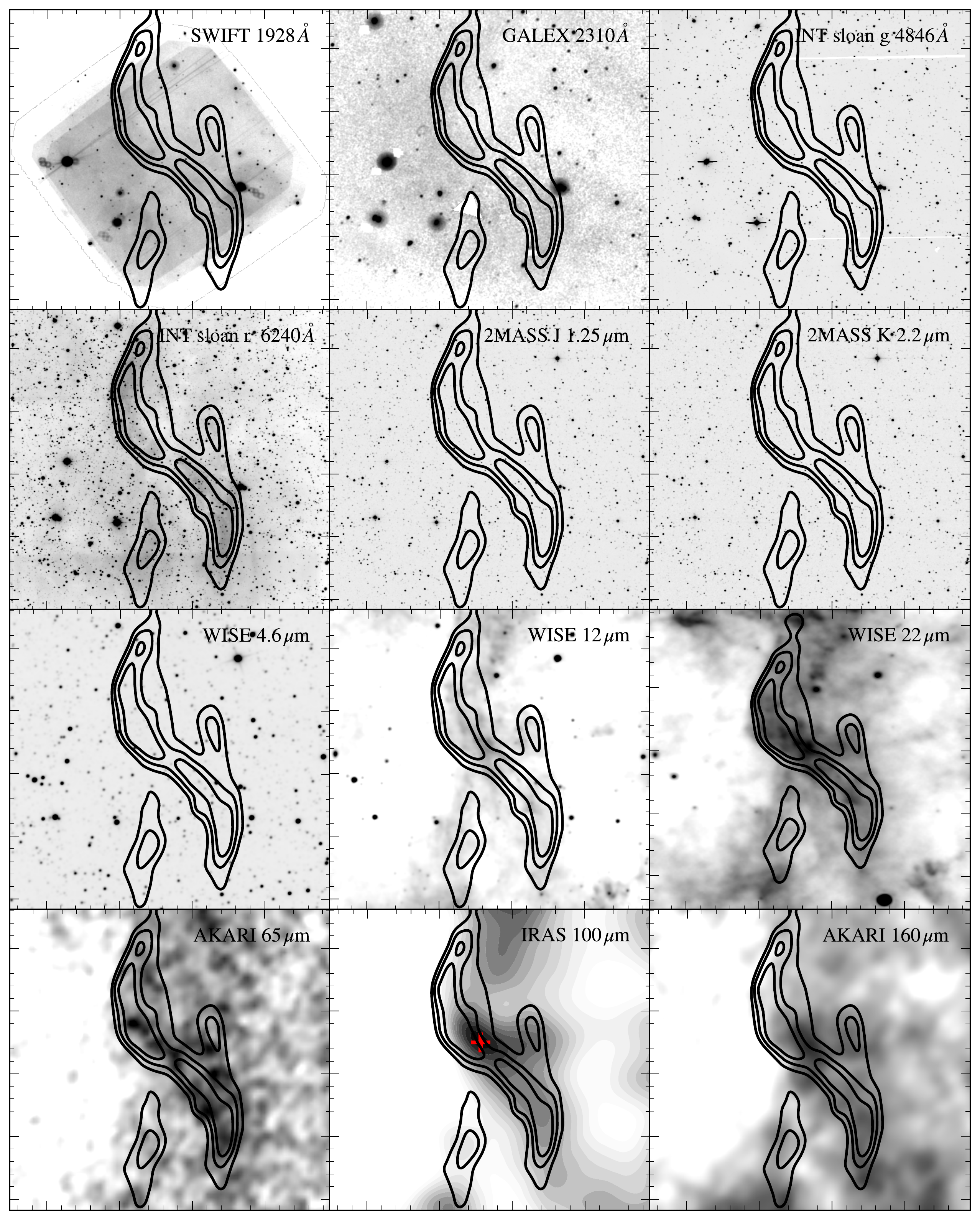}
\centering
\caption{UV, optical, and infrared images of Simeis 57.  Each image
  shows the same region of $0.4\times 0.4\,\mathrm{deg}$ centered on
  $\alpha=304.04$, $\delta=43.69$. The UV images comprise SWIFT UVOT
  ($\lambda_{eff} = 1928 \, \mathrm{\AA}$) and GALEX near-UV
  ($\lambda_{eff} = 2310 \, \mathrm{\AA}$) exposures.  Optical
  broadband images include those from our INT WFC observations with
  the filters sloan $g$ ($\lambda_{0} = 4846 \, \mathrm{\AA}$) and
  sloan $r$ ($\lambda_{0} = 6240 \, \mathrm{\AA}$).  Near-infrared
  images are 2MASS $J$ and $K_s$ band at wavelengths of 1.25
  $\mathrm{\mu m}$ and 2.2 $\mathrm{\mu m}$ and WISE band 2 to 4
  images at wavelengths of 4.6, 12, and 22
  $\mathrm{\mu m}$.  Far-infrared images are AKARI 65, enhanced IRAS
  100, and AKARI 160 $\mathrm{\mu m}$.  The point source IRAS
  20145+4333 is denoted by a red cross in the HIRES
  $100 \mathrm{\mu m}$ map. Contours correspond to DRAO 1420MHz radio
  continuum intensities of 20, 25, 30 $\mathrm{mJy \, arcmin^{-2}}$.}
\label{fig:image_spectrum}
\end{figure*}

% Figure 6

\begin{figure*}
\includegraphics[width=1\textwidth]{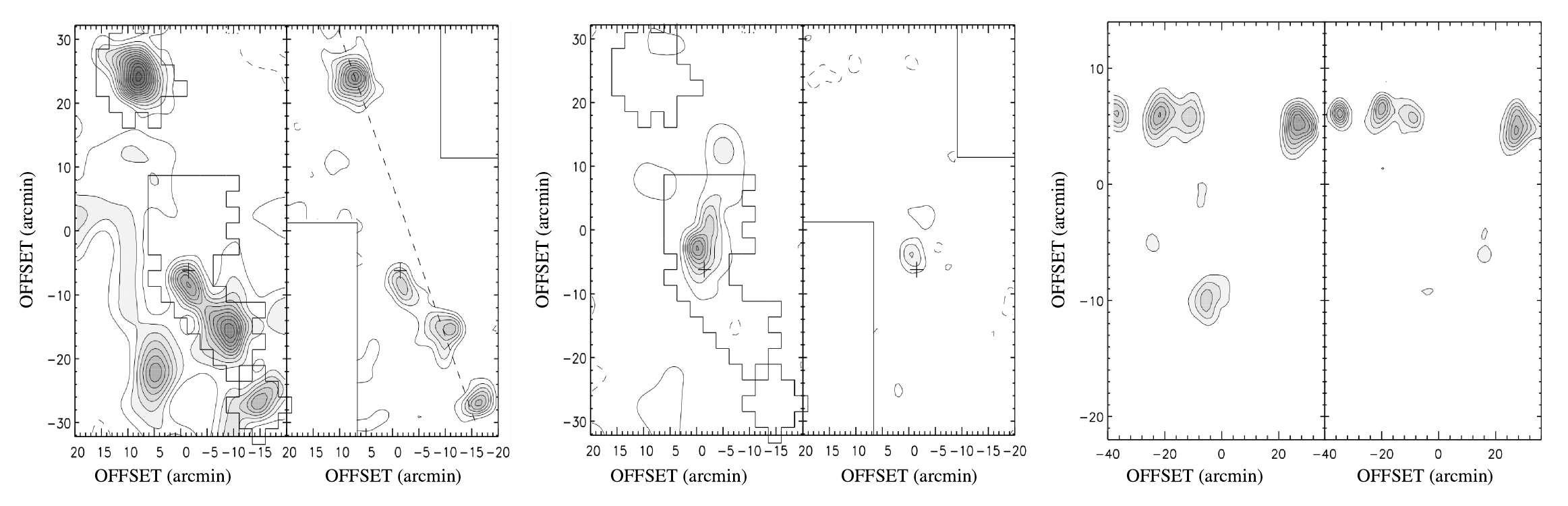}
\caption{Undersampled $J$=1-0 $^{12}$CO (left panels) and $^{13}$CO
  (right panels) maps of molecular clouds in the field of Simeis~57,
  whose center is denoted by a cross. The emission at +5 km s$^{-1}$ is
  shown in the leftmost two panels (averaged over a 5 km s$^{-1}$
  velocity interval), and the central two panels show the emission at
  -10 km s$^{-1}$ (averaged over a 4 km s$^{-1}$ interval). In the
  $^{12}$CO panels, the regions with improved sampling are indicated
  by a solid-line boundary.  The rightmost two panels show the
  position-velocity map along the dashed line in the $^{13}$CO
  map. Position offsets are relative to $\alpha\,=\,20^h16^m16.7^s$,
  $\delta\,=\,+43^{\circ}47'54"$ (J=2000)}
\label{fig:blt_co_maps}
\end{figure*}

\subsection{UV data: Swift and GALEX}

% Figure 7: JCMT CO maps

\begin{figure*}
\includegraphics[width=1\textwidth]{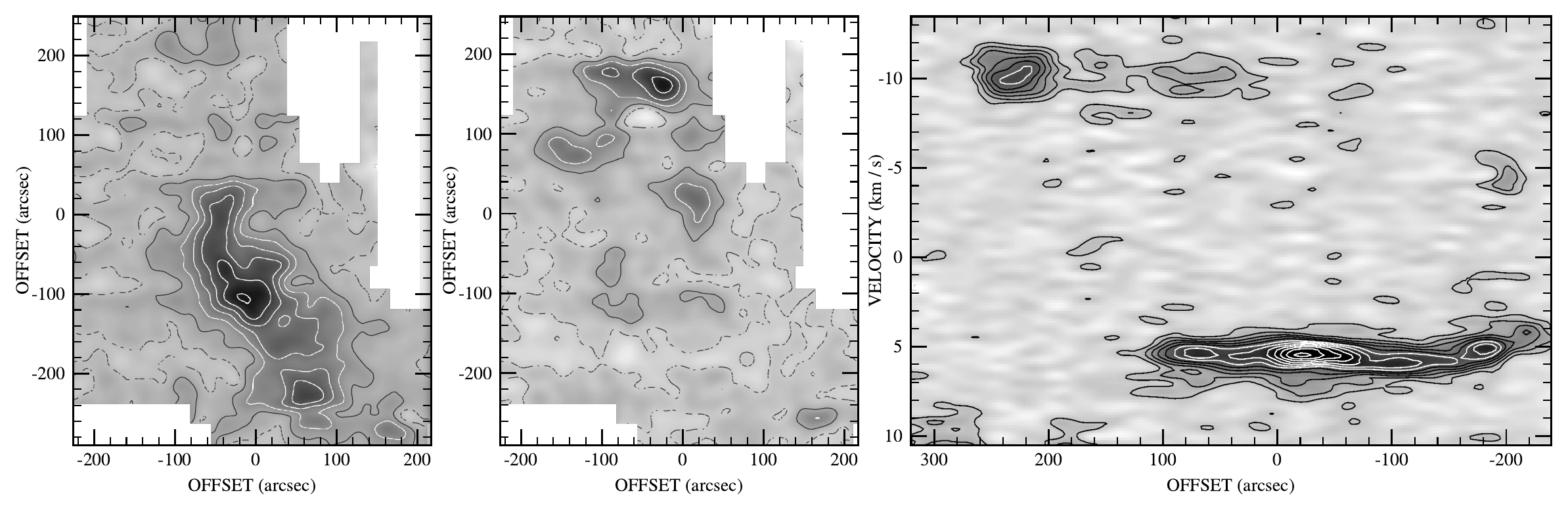}
\centering
\caption{Fully sampled $J$=2-1 $^{13}$CO maps in the field centered on
  Simeis~57 (offset relative to $\alpha=20^h16^m08.2^s$,
  $\delta=43^{\circ}41'02.4''$) smoothed to a resolution of $30"$. At
  left is the emission at +5 km s$^{-1}$ (averaged over a 6 km
  s$^{-1}$ velocity interval), in the center the emission at -10 km
  s$^{-1}$ (averaged over a 10 km s$^{-1}$ interval), and at right the
  position-velocity distribution.}
\label{fig:jcmt_co_maps}
\end{figure*}

We obtained a new UV image of the Simeis~57 field with the NASA Neil
Gehrels Swift Observatory, a multiwavelength space observatory
launched in 2004. The onboard Ultraviolet/Optical Telescope (UVOT) is
a modified Ritchey-Chr\'etien telescope with an aperture of 30 cm and
an f/2.0 primary reimaged to f/13. Its 2048$\times$2048 pixel
  detection array combines a pixel scale of $0.5"$ with a
$17'\times17'$ field of view \citep{2008MNRAS.383..627P}. The
observations were made between 2012 March and 2013 January in the UVW2
filter, which has a maximum efficiency at
$\lambda_{center} = 2000\mathrm{\AA}$.  At this wavelength, the
resolution is about $1.5"$.

\par We also retrieved near-UV
($\lambda_{eff} = 2310 \, \mathrm{\AA}$) observations of the region
surrounding DWB~111 from the archive of the NASA GALEX space
observatory (2003-2012). The GALEX observatory is equipped with 
a 50 cm aperture telescope
with a field of view of 1.2 degrees wide. We used MOSAIX
\citep{armengot2014mosaix} to stack the retrieved images.  In
Fig.\,\ref{fig:image_spectrum} we show both UV images.

\subsection{Optical data: Gaia}

Gaia \citep{refId0} is an European Space Agency (ESA) space observatory 
that was launched in 2013. It uses two telescopes to perform astrometry,
photometry, and spectrometry of stars with very high precision down to
a visual magnitude of 20. In the absence of visual extinction, it
should thus see OB stars across the entire Galaxy. It was designed to
provide position, parallax, and annual proper motion with accuracies
of $0.025$ mas at $V$ -15 mag to $0.6$ mas at $V$=20 mag, and
distances accurate to $\leq1\%$, and $\leq10\%$, respectively.

\par From the second data release (DR2) in 2018, we extracted 218136
stars within a radius of one degree from the nebula and selected the
13381 stars ($6\%$) with high-quality data (RUWE $<$ 1.40) shown in
Fig.\,\ref{fig:gaia_figs} \footnote{We verified that the improvements
  in the more recent Gaia eDR3 do not discernibly change
  the results described in this work}. Because stars associated with the
ionization of Simeis 57 should be hot, we specifically
  considered the 49 bright, blue stars with $M_{\rm G}\,<\,3$ and
$G_{\rm BP}-G_{\rm RP}\,<\,0.4$. This sample defines the peak of the
main-sequence region in the Hertzsprung-Russell (HR) diagram of the Simeis 57 region.
Many of these stars are visible in the NASA GALEX and ESO Digitized Sky Survey
(DSS2) data, and this includes a few close to the propeller nebula itself.

\subsection{Infrared data: IRAS, 2MASS, WISE, and Akari}

We used the public databases kept at the NASA/IPAC
Infrared Science Archive (IRSA) to investigate the infrared properties of 
Simeis~57 and surroundings. Far-infrared maps were constructed
from the {\it IRAS} and the {\it Akari} data, near- and mid-infrared
maps from the {\it WISE} data, and near-infrared maps from the {\it
  2MASS} data.

\par We used enhanced resolution {\it IRAS} \citep{Neugebauer_IRAS}
infrared continuum maps at 12, 25, 60, and 100 $\mathrm{\mu m}$ from
the Infrared Processing and Analysis Center (IPAC). The  HIRES
  enhancements employs the maximum correlation method (MCM)
  \citep{Aumann1990} in 20 iterations to construct
resolution-enhanced coadded {\it IRAS} images. The original {\it IRAS}
resolutions vary between $0.5'$ at 12 $\mathrm{\mu m}$ to about $2'$
at 100 $\mathrm{\mu m}$. The HIRES enhancements provided a roughly 
five-fold increase
in resolution, but the actual resolution varies over the map.
A tilted plane was fitted to the background and subtracted. The
{\it IRAS} map in Fig.\,\ref{fig:image_spectrum} is dominated by
diffuse extended emission from warm dust.

\par
Between 1997 and 2001, the University of Massachussetts carried
out an all-sky survey in the $J$ 1.25 $\mathrm{\mu m}$, $H$ 
1.65 $\mathrm{\mu m}$, and $K_{s}$ 2.17 $\mathrm{\mu m}$ bands
\citep{skrutskie2006two}. The resulting digital {\it 2MASS} sky atlas
has a resolution of $4"$ in each of the three bands.
In Fig.\,\ref{fig:image_spectrum} we show the $J$ and $K_{S}$ band
images centered on Simeis~57. These images do not show emission from
either nebula or dust but are dominated by the stars in the field.

\par We downloaded NASA Wide-Field Infrared Survey Explorer ({\it WISE}, 
2009-2011) archival images with a field of view of $47'$ and a
resolution of $6"$ in four bands at 3.4, 4.6, 12, and
22$\mathrm{\mu m}$. The last two have a much higher sensitivity
than the corresponding {\it IRAS} bands.
In Fig.\,\ref{fig:image_spectrum}, the near-infrared (4.6
$\mathrm{\mu m}$) image is dominated by the stars in the field, but
the two mid-infrared (12 $\mathrm{\mu m}$ and 22 $\mathrm{\mu m}$)
images show emission from hot dust associated with the nebula. 
Unlike the {\it IRAS} images, these very sensitive {\it WISE}
images still show the brightest stars.

\par The {\it Akari} infrared observatory, developed by the Japan
Aerospace Exploration Agency (JAXA) with ESA participation,
conducted an all-sky survey (2006-2007) in four F bands
at 65, 90, 140, and 160 $\mathrm{\mu m}$. This instrument has a resolution of
$1'$ to $1.5'$ \citep{AkariPaper}
comparable to that of HIRES, with the advantage of being constant
across the field. The data were released in 2014 as
$6^{\circ}\times6^{\circ}$ FITS images covering the whole sky. 
Like the {\it IRAS} images, the {\it Akari} images at 65 and 160
$\mathrm{\mu m}$ in Fig\,\ref{fig:image_spectrum}
are dominated by the diffuse thermal emission from warm dust.

\subsection{Millimeterwave CO data: BTL, NRAO, and JCMT}

Dense clouds of dust and gas can hide luminous stars from view
and Simeis~57 may be associated with a potentially much larger
molecular cloud complex not seen at optical or infrared wavelengths.
We used the Bell Telephone Laboratories (BTL) 7 m telescope in New
Jersey in 1983 to observe the $J$=1-0 $^{12}$ CO (115 GHz) and the
$^{13}$ CO (110 GHz) transitions in frequency-switched mode with 
a clean Gaussian beam of 100$"$. The observing and reduction
techniques were identical to those described by \cite{Bally1987}.  The
undersampled $40'\times64'$ maps reveal five small clouds with
$V_{LSR}\,\sim\,+5$ km s$^{-1}$.  In four of these clouds we increased the map
sampling to a $2.5'$ grid.
At the map center, a fifth cloud occurs at a separate velocity
$V_{LSR}$ = -10 km s$^{-1}$ (Fig.\,\ref{fig:blt_co_maps}). Additional
$^{12}$CO and $^{13}$CO observations, made in 1989 with the NRAO 12 m
telescope,
yielded higher-resolution ($55"$) $^{12}$CO/$^{13}$CO
isotopolog ratios at various positions in the central cloud.
Finally, the central cloud was mapped in the J=2-1 transition with the
James Clerk Maxwell Telescope (JCMT) in various observing runs between
2002 and 2005 at an even higher resolution of $22"$. The fully sampled
($5"$ - $10"$ grid) $^{13}$CO map covers an area $420"\times560"$
(Fig.\,\ref{fig:jcmt_co_maps}). In $^{12}$CO two smaller
maps were obtained centered on offsets (0, +200$"$) and (-80$"$,
-30$"$), respectively.

\section{Analysis}

\subsection{Distance}

\begin{figure}
\includegraphics[width=0.48\textwidth]{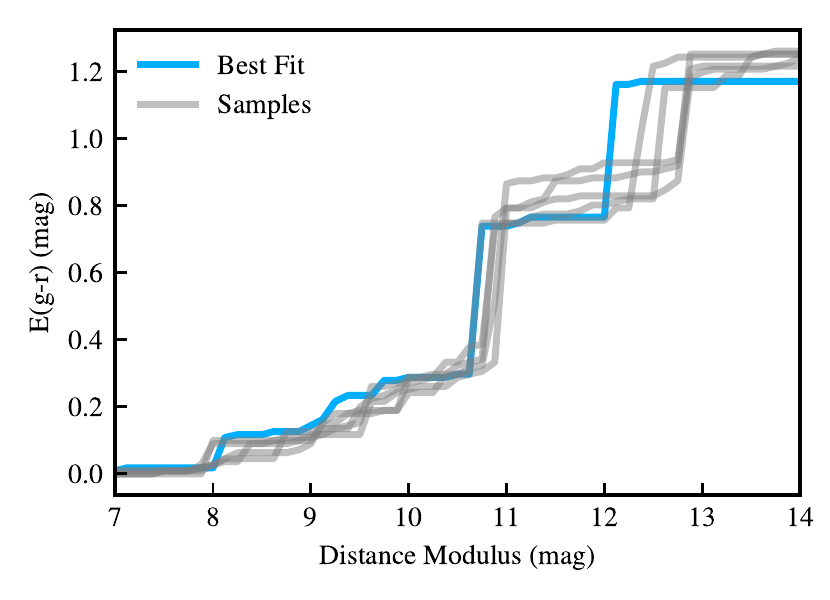}
\centering
\caption{Reddening with distance based on \cite{2019arXiv190502734G}.
  The gray lines indicate stellar samples; the blue line indicates the best
  model. }
\label{fig:reddening_with_D}
\end{figure}

\cite{israel2003} suggested that Simeis~57 is relatively nearby
and found the extinction in front of the nebula to vary between
$A_V$ = 1.0$^{m}$ and $A_V$ = 2.8$^{m}$ with a mean of 2$^{m}$.
Recent work based on Gaia, Pan-STARRS 1, and 2MASS by
\cite{2019arXiv190502734G}
provides more detail on the reddening E(g-r)\footnote{For an $R_{V}$ =
  3.1 reddening law, $E(B-V)\,\sim0.981\,E(g-r)$
  \citep[see][]{Schlafly2011}.} of field stars in the direction of
Simeis~57 as a function of distance. 
In Fig.\,\ref{fig:reddening_with_D}, steep increases in reddening indicate
the presence of dust sheets at $m- M = 10.5^{m}$ ) and $m- M= 12^{m}$,
in which extinction jumps from $A_V$ = 1.0$^{m}$ to
2.4$^{m}$ at $1.3\pm0.1$ kpc and from $A_V$ = 2.5$^{m}$ to 3.6$^{m}$
at $2.5\pm0.2$ kpc; this suggests that Simeis~57 is located between
these distances.

\par
\begin{figure}
\includegraphics[width=0.48\textwidth]{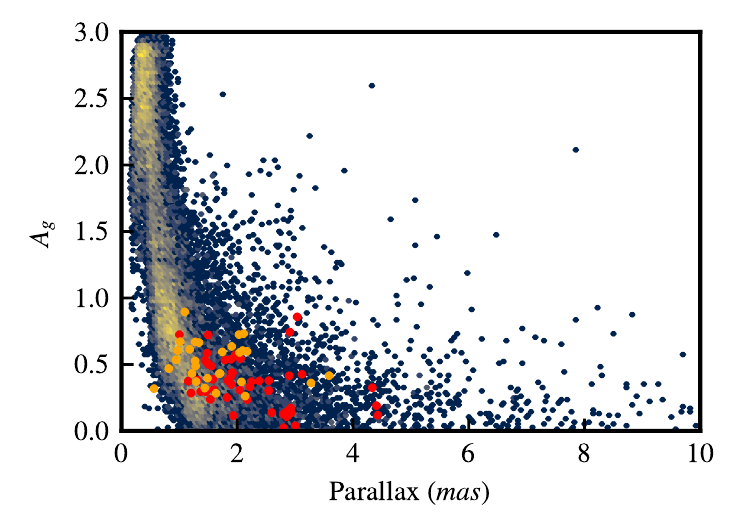}
\centering
\caption{Gaia visual extinction ($A_g$ in mag) vs. parallax
  (in milli-arcsec) for all stars with high-quality data within one
  degree of Simeis~57. Highlighted in this sample are 49 bright blue
  stars moving in predominantly NS (yellow dots) or EW (red dots)
  directions.}
\label{fig:gaia_figs}
\end{figure}

 The distribution of extinction and parallaxes of the Gaia sample
stars is shown in Fig.\,\ref{fig:gaia_figs}, in which colored dots 
denote the bright blue stars mentioned earlier. The parallax distribution
has a sharp edge at about 0.35 milli-arcsec.  This corresponds to a
distance of about 3 kpc and a height above the plane of 250 pc. At
that point the line of sight clears the Milky Way disk so that 3 kpc
is the upper limit to the distance of Simeis~57. The subset of bright
blue stars samples a relatively large range of distances between 350
pc and 1500 pc, but it is not evident that these stars are associated
with each other or that Simeis~57 is related to any of them.
Although the evidence is still inconclusive, we adopt a
distance of $D\,=\,1.7\pm0.5$ kpc, which we substantiate in more
detail in Sect. \ref{section_starsinfield}.

% Figure 8: Reddening vs distance modulus

%Figure 9: Gaia Figures

\subsection{The nature of the nebular gas}

% Table 2: Spectral line ratios
\begin{table}[ht]
{\small %
  \caption{Diagnostic spectroscopic line ratios$^a$} 
  \label{tab:spectral_line_ratios}
\centering
\begin{tabular}{p{0.24\linewidth}p{0.14\linewidth}p{0.14\linewidth}p{0.14\linewidth}p{0.14\linewidth}}
\toprule
                    & North 2 & North 1 & Center & South \\
\midrule
F(H$\alpha$)/F([NII])$^b$ &  $2.2\pm0.08$ &  $2.0\pm0.19$ &  $1.9\pm0.21$ &  $1.7\pm0.12$ \\

F(H$\alpha$)/F([SII])$^b$ &  $4.1\pm0.12$ &  $3.8\pm0.14$ &  $3.2\pm0.27$ &  $2.5\pm0.11$ \\

F(H$\alpha$)/F(H$\beta)$  &  $4.9\pm0.29$ &     $...$      &     $...$      &  $5.8\pm0.51$ \\

%[NII]6548/6584      &  $3.1\pm0.24$ &  $2.8\pm0.53$ &   $3.0\pm0.7$ &  $3.0\pm0.48$ \\

%[NII] 6584/H\alpha &  $0.3\pm0.01$ &  $0.4\pm0.02$ &  $0.4\pm0.04$ &  $0.4\pm0.02$ \\

F([OIII])/F(H$\beta$)     &  $0.5\pm0.03$ &     $...$      &     $...$      &  $0.1\pm0.05$ \\

F([SII]6716/6731)      &  $1.4\pm0.06$ &  $1.5\pm0.11$ &  $1.4\pm0.21$ &  $1.4\pm0.12$ \\

\bottomrule
\end{tabular}
\\
Note: a. Values indicate observed ratios for the positions denoted in Fig.\,\ref{fig:Ha_overview}; 
spectra shown in Fig.\,\ref{fig:spectra_3incol}.
b. [SII] and [NII] indicate the sum of the doublets.
}
\end{table}

\par In all spectra shown in Fig.\,\ref{fig:spectra_3incol}
the [NII] and [SII] doublets are well resolved, and
the relatively weak [OI]6300\AA\, and [OI]6334\AA\, lines can be made out.
The H$\beta$4861\AA\, and [OIII]5007\AA\, are seen only in the S and N2
spectra, which extend sufficiently blueward (recall Fig. \ref{fig:spectra_3incol}).

A comparison of the diagnostic line ratios in
Table\,\ref{tab:spectral_line_ratios} with those of the Galactic HII
region IC~1805 \citep[Figs. 12 and 13 in][]{Lagrois2012} clearly
shows that the Simeis~57 spectra are those of photoionized gas in an HII
region at all positions. Only at the southern position a small
amount of shock-ionized gas may be mixed in as well. The [SII]
doublet ratios of about 1.4 indicate low
electron densities $n_e\sim 30\, \mathrm{cm^{-3}}$ to
$n_e\sim 100\, \mathrm{cm^{-3}}$ if $T_e=10000\, \mathrm{K}$
\citep{2006agna.book.....O}. These
are very close to the r.m.s. electron densities $<n_e^2>^{1/2}$ of
$60\pm15\, \mathrm{cm^{-3}}$ determined from the radio maps by
\cite{israel2003} adapted to $D=1.7$ kpc, implying clumping factors of
about unity, that is, a smooth ionized gas distribution.

%Table 3: Line fluxes
\begin{table}[h]
{\small %
  \caption{Area-integrated line fluxes}
  \label{tab:fluxes_info_from_maps}
\begin{center}
\begin{tabular}{lcccc}
\toprule
Region$^a$ & \multicolumn{4}{c}{Flux$^b$ ($10^{-11}\, \mathrm{erg\, cm^{-2}\,s^{-1})}$} \\
    &       [SII]  & H$\alpha^c$  &    H$\beta$  &  [OIII]        \\
\midrule
A   &   8.6 $\pm$ 2.1 &   30.8 $\pm$ 7.7 &  5.6 $\pm$ 1.4 &  1.7 $\pm$ 0.4 \\
A+B &  17.4 $\pm$ 4.3 &   57.9 $\pm$14.5 &  9.9 $\pm$ 2.5 &  2.5 $\pm$ 0.6 \\
B   &   8.8 $\pm$ 2.2 &   27.1 $\pm$ 6.8 &  4.3 $\pm$ 1.1 &  0.9 $\pm$ 0.2 \\
C   &   1.6 $\pm$ 0.4 &    5.2 $\pm$ 1.3 &  0.9 $\pm$ 0.2 &  0.2 $\pm$ 0.1 \\
D   &   3.2 $\pm$ 0.8 &   11.3 $\pm$ 2.8 &  2.0 $\pm$ 0.5 &  0.8 $\pm$ 0.2 \\
\bottomrule
  \end{tabular}
  \end{center}
Notes: a. Defined in Fig.\,\ref{fig:Ha_overview};
b. Continuum-subtracted observed fluxes with residual stellar emission masked;
c. Corrected for a $\sim35\%$ contribution by [NII] as suggested by
Fig.\,\ref{fig:spectra_3incol} and Table 2.} 
\end{table}

\par
\begin{figure}
   \includegraphics[width=0.48\textwidth]{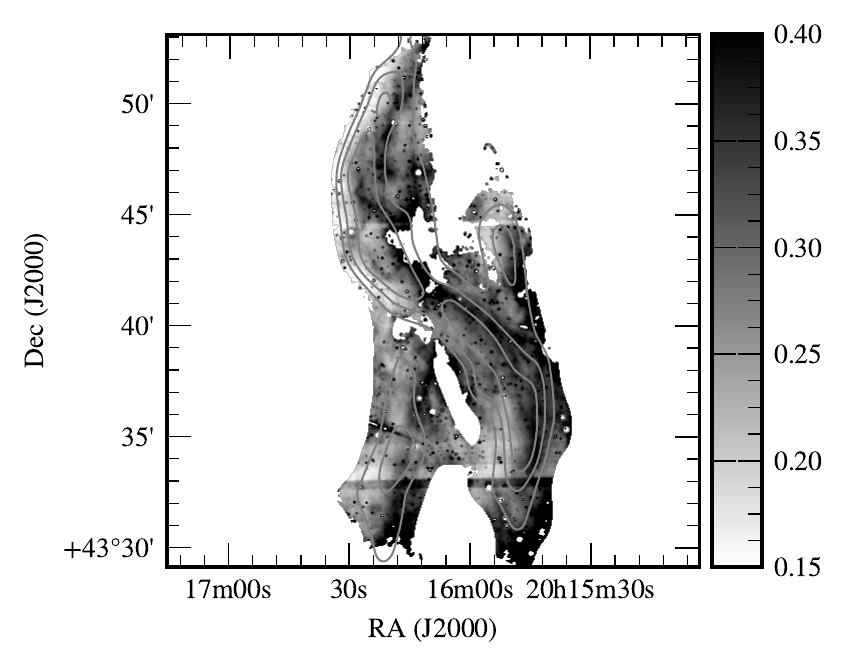}
 \centering
 \caption{
   Distribution of [SII]/H$\alpha$ line flux ratios. Only
   pixels with significant emission are included. The observed
   H$\alpha$ intensities have been corrected for a $35\%$ contribution
   from the nearby [NII] lines. The dark spots, gaps, and a horizontal bar
   are artifacts from different seeing conditions and the dithering
   procedure. The light gray contours are radio contours as in Fig. 
   \ref{fig:Ha_overview}.}
\label{fig:SII_over_Ha}
\end{figure}

 The H$\alpha$, H$\beta$, and [SII] images in
Fig.\,\ref{fig:line_images_4onrow} reveal very similar spatial
distributions of ionized hydrogen and ionized sulfur, differing only
in detail.
The peculiar ``propeller'' shape of the nebula stands out; the clear
break in the middle is apparently caused by intervening dust.
Taking our cue from the line spectra in Fig.\,\ref{fig:spectra_3incol},
we subtracted a constant [NII] contribution of $35\%$ from the
emission measured in the H$\alpha$ filter before constructing the
[SII]-to-H$\alpha$ ratio map in Fig\,\ref{fig:SII_over_Ha}.
Over most of the nebula, this ratio is $\sim0.35$, as expected for thermal
emission from HII regions \citep[cf. Fig. 12 from][]{Lagrois2012}. Closer 
inspection of Fig.\,\ref{fig:SII_over_Ha}, however, reveals a thin ridge of ratios
$\geq 0.4$ along the western edges of both DWB~111 and DWB~119, that is,
higher than expected for pure thermal emission. This is marginal
evidence for local shocks occurring at the edge of Simeis~57,
which is consistent with the comment we made above. 
The radio continuum spectral index map
\citep[Fig. 7 of][]{israel2003} does not show any trace of
nonthermal radio emission so that this is at best a minor
constituent.

% Figure 10: [SII] over Ha with radio contours

\par [SII] emission is prominent in region C (the ``spur'') northwest of
the obscured center where no oxygen is seen.  The intensities of
ionized sulfur [SII] and ionized oxygen [OIII] are largely
anticorrelated as essentially all sulfur is ionized up to [SIII] or
[SIV] in the [OIII] emitting zone.  The distribution of ionized oxygen
([OIII]) is very different from that of the ionized hydrogen.  It is
more diffuse and the propeller shape is not prominent.  In region A
(DWB~111), the [OIII] emission is in thin ridges east of the northern
propeller ``blade''.  There is virtually no diffuse [OIII] emission west
of it.

The southern propeller region B (DWB~119), which is bright in the
Balmer lines and the radio continuum, does not show up at all in
[OIII]. There is, however, relatively bright and extended [OIII]
emission to its east, also in directions of weak Balmer line emission.

\par The {\it WISE} mid-infrared images and especially the {\it IRAS}
and {\it Akari} far-infrared images likewise show dust and [OIII]
emission to be anticorrelated.
They show diffuse extended infrared emission west of the propeller but
a near lack east of the nebula, where the [OIII] emission is
prevalent. It appears that there is a general lack of material in
sight lines east of Simeis~57 and that the ionized gas at the eastern
edges is much more highly excited than the dusty gas to the west.

\subsection{Dust extinction toward Simeis~57}
\par To establish properties of Simeis~57 such as its
H$\alpha$ luminosity, distance, and source of ionization, we need to
constrain the dust extinction toward Simeis~57. This can be done
using the Balmer line and radio continuum intensities. The H$\alpha$ 
and H$\beta$ line intensities and the color excess are related 
via \citep[see for details][]{Momcheva_2013}
\begin{equation}
\begin{split}
    E(B-V) &= \frac{E(H\beta-H\alpha)}{k(\lambda_{H\beta})-k(\lambda_{H\alpha})}\\
    &= \frac{2.5}{k(\lambda_{H\beta})-k(\lambda_{H\alpha})} \log \Big[\frac{(H\alpha/H\beta)_{int}}{(H\alpha/H\beta)_{obs}} \Big]
\end{split}
\end{equation}
\noindent
where $k(\lambda$) is the wavelength-dependent value of the reddening curve and
$(H\alpha/H\beta)_{int}$ and $(H\alpha/H\beta)_{obs}$ are the
unreddened and observed line ratios, respectively.
Assuming Case B recombination at $T_e=10000\,\mathrm{K}$ conditions
\citep{2011piim.book.....D}, we expect (H$\alpha$/H$\beta$)$_{int}=$
2.86.  $E(H\beta-H\alpha)$ is defined analogous to $E(B-V)$.
For a standard Milky Way reddening law \citep{Cardelli1989} we find
\begin{equation}
\begin{split}
    E(H\beta - H\alpha) &= -2.50 \log \Big[\frac{(H\alpha/H\beta)_{int}}{(H\alpha/H\beta)_{obs}} \Big] = 0.77\pm0.37\\
    E(B-V) &= -1.97 \log \Big[\frac{(H\alpha/H\beta)_{int}}{(H\alpha/H\beta)_{obs}} \Big] = 0.60\pm 0.29
\end{split}
\end{equation}
With
\begin{equation}
    A_{\lambda} = k(\lambda) E(B-V)
    \label{eqk_lambda}
\end{equation}
we obtain
\begin{equation}
\begin{split}
    A_V \,&= 3.1\times E(B-V)=1.9\pm0.9\,\mathrm{mag}\\
    A_{H\alpha} &= 2.6\times E(B-V) = 1.6\pm0.8\,\mathrm{mag}.    
\end{split}
\end{equation} 
The extinction implied by the reddening of the Balmer lines is
consistent with the result obtained by \cite{israel2003} and the
average Galactic visual extinction rate of 1.8$^{m}$ per kiloparsec
\citep{whittet2002dust}.

\par The H$\alpha$ line and 1420 MHz radio continuum images are very
similar since the emission mechanisms are intimately related.  Unlike
the H$\alpha$ line, the emission at radio frequencies does not suffer
extinction and the intensity ratio of the two images allows for a direct
determination of the extinction across the entire nebula.
With the new H$\alpha$ image, the earlier attempt by
\citep{israel2003} can be improved in sensitivity and detail;
the limiting factor is now the resolution ($58"\times 80"$) of the 1420
MHz DRAO radio continuum map.  For electron temperatures
$T_e=10 000 \, \mathrm{K}$, we expect optically thin free-free
continuum and H$\alpha$ line emissivities
\begin{equation}
\begin{split}
    \epsilon_{1420} &= 3.9 \times 10^{-39} \, \mathrm{n_e^2 \, erg \, s^{-1} \, cm^{-3}\, Hz^{-1}}\\
    \epsilon_{H\alpha} &= 3.6 \times 10^{-25} \, \mathrm{n_e^2 \, erg \, s^{-1} \, cm^{-3}},
\end{split}
    \label{eq:opticallythin1420}
  \end{equation}

yielding an extinction-free ratio

\begin{equation}
 \frac{S_{1420}}{F(H\alpha)} = 1.15\times 10^{-14}\, \mathrm{Hz^{-1}}
\label{eq:relation_radio_Ha}
\end{equation}

\noindent which may be compared to observed ratios. In the extinction
map thus constructed (Fig.\,\ref{fig:co_extinction_map}), we
masked the regions with $F_{1420}\leq 10\, \mathrm{mJy \,arcmin^{-2}}$
and assumed an extinction law $A(\lambda)\sim \lambda^{-1}$, so that
$A_V \approx 1.2 A_{H\alpha}$.
\par In this map, the elongated extinction feature north of the
center of Simeis~57 coincides with part of the long radio filament
mentioned before. Its southern continuation has a radio counterpart
(DBW~118; region D) seemingly suffering only modest extinction. Most
of the bright H$\alpha$ emission associated with the
radio source suffers relatively little extinction and is consistent
with dust in the foreground line of sight. The extinction in
Fig.\,\ref{fig:co_extinction_map} agrees well with the reddening
implied by the Balmer lines alone, leaving little room for the
``gray'' extinction that would betray dust internal to the ionized
gas.  The strong absorption feature
just to the northeast of the nebula center has an infrared
counterpart cataloged as IRAS PSC 20145+4333, denoted in
Fig.\,\ref{fig:image_spectrum}, which appears diffuse and resolved.

\par We used the mean extinction $A_{H\alpha} $= 1.6$^{m}$ to
convert the H$\alpha$ flux of region A+B in Table
\ref{tab:fluxes_info_from_maps} into the intrinsic H$\alpha$
luminosity

\begin{equation}
\begin{split}
    L_{H\alpha} &= 4\pi D^2 F_{H\alpha} 
    10^{0.4 A_{H\alpha}} \\
    &= (9\pm 2)\times 10^{35} \left(\frac{D}{1700\, pc}\right)^2\, 
    \mathrm{erg \, s^{-1}}
\end{split}
    \label{eq:Hafluxtoluminos}
\end{equation}
For case B recombination, $T=10 000 \, \mathrm{K}$ and
$n\sim 10^2\,\mathrm{cm^{-3}}$, the number of absorbed (stellar)
Lyman continuum photons $N_{Lyc}$ is related to $L_{H\alpha}$ via
\citep[cf.][]{Kennicutt:1994wu}

\begin{equation}
\begin{split}
    N_{Lyc} &=  7.1 \times 10^{11} L(H\alpha)\\
    &=(6\pm1.5)\times 10^{47}\, \left(\frac{D}{1700\, pc}\right)^2\, \mathrm{s^{-1}}.
    \label{eq:LyafromHa}
\end{split}
\end{equation}

\subsection{Infrared emission from dust}

\par In Fig.\,\ref{fig:image_spectrum}, emission from hot dust is
seen at $12\,\mathrm{\mu\,m}$ and $12\,\mathrm{\mu\,m}$ tracing
the linear filament separating regions A and B. It is unrelated to
the nebula itself and has no far-infrared counterpart from cooler
dust.

The propeller-shaped nebula is seen at wavelengths from
$22\,\mathrm{\mu\, m}$ to $100\,\mathrm{\mu\, m}$.  In the WISE
$22\,\mathrm{\mu\, m}$ image, the emission from hot dust traces the
northern DWB~111 blade, whereas warm dust in the southern DWB~119
blade is more prominent in the Akari $65 \mu$m image . The compact
dusty region IRAS 20145+4333 (see IRAS $100\,\mathrm{\mu\, m}$ image)
is clearly present longward of $22\,\mathrm{\mu\, m}$. It does not
have the appearance of an embedded stellar object. In comparison with
other Galactic sources, the infrared emission from Simeis~57 is
insignificant. Unlike its optical counterpart, the object does not
stand out in infrared images at any wavelength.

% Table 4 Infrared luminosities

\begin{table}
\caption{Area-integrated infrared fluxes.}
   \label{table:IR_luminosities}
\begin{center}
\begin{tabular}{lrrcc}
\toprule
Mission& $\lambda$ & $\nu$ & Flux$^{a}$   & Flux$^{b}$    \\
       &($\mu$m)   & (THz) & (Jy)        & (Jy)          \\
\midrule
IRAS   &        12 &    25 &  13$\pm$4   &   56$\pm$42   \\
IRAS   &        25 &    12 &  35$\pm$7   &  136$\pm$83   \\
IRAS   &        60 &     5 & 326$\pm$67  & 1350$\pm$781  \\
AKARI  &        60 &     5 & 275$\pm$67  & 1265$\pm$657  \\
AKARI  &        90 &     3 & 251$\pm$18  & 1259$\pm$578  \\
IRAS   &       100 &     3 & 511$\pm$100 & 2368$\pm$1265 \\
AKARI  &       140 &     2 & 173$\pm$41  &  807$\pm$431  \\
\bottomrule
\end{tabular}
\end{center}
Notes: a. PSC IRAS 20145+4333, integrated over a solid angle of
$76\, \mathrm{arcmin^2}$.  b. Entire nebula, integrated over
a solid angle of $550\, \mathrm{arcmin^2}$.
\end{table}

\par We extracted the infrared brightnesses
% at the frequencies
available from the {\it IRAS} and {\it Akari} databases, estimating
the infrared background from the empty region east of the object. We
derived the area-integrated infrared fluxes of the central
compact object IRAS 20145+4333 and the whole nebula
(Table\,\ref{table:IR_luminosities}).  The infrared luminosity of the
compact region is only
$L_{\rm TIR}(compact)\,\approx\,1.9\times    10^{3}\times(D/1700
  pc)^2\, \mathrm{L_{\odot}}$ \citep[using][]{Lee_1996}.  Its
spectral shape matches that of an embedded OB star
\citep{wood1989massive} but the luminosity is orders of magnitude
below that expected for such a star. Even when integrated over the
much larger area of $550\, \mathrm{arcmin^{2}}$, the infrared
luminosity of the entire nebula remains low at
$L_{\rm TIR}(Si57)\,\approx 9\times 10^{3}\times(D/1700
pc)^2\,\mathrm{L_{\odot}}$. By fitting a modified blackbody
\citep{battersby2011characterizing} to the AKARI and IRAS data with a
$\beta=1.6$ emissivity, we find a model dust temperature
$T_D \approx 36\, \mathrm{K}$.

At the assumed distance of $1.7$ kpc, the above value of $L_{\rm TIR}$
corresponds to the luminosity of an embedded $\sim$B1.5V star with an 
absolute visual magnitude $M_V\,\approx\,-2.8\, \mathrm{mag}$
\citep{1973AJ.....78..929P,Crowther2005}.  This
is only a minimum estimate as the radiating dust may subtend a solid
angle as seen from the star substantially less than $4\pi$.  For
instance, if only $10\%$ of the sky were blocked by dust, an O7 star
with $M_V\,=\,-4.2\, \mathrm{mag}$ would be needed, and with a
blockage of only $1\%$ an O4/O5 star with
$M_V\,=\,-5.8\, \mathrm{mag}$.
% would be required.
The analysis of the observed radio continuum or Balmer line emission
leads to similar results.

\subsection{Molecular clouds and Simeis~57}

The observations show surprisingly little CO emission toward Simeis
57. Unlike typical HII regions, the optical nebula is much larger
  than the associated molecular cloud. The surrounding field lacks
  major molecular clouds and only contains a number of isolated small
molecular clouds in a narrow velocity range ($V_{LSR}$ = +4 to +6 km
$^{-1}$) Fig.\,\ref{fig:blt_co_maps}. These ``red'' clouds,
of 3 to 8 pc in size, coincide with undistinguished far-infrared features
(Fig.\,\ref{fig:image_spectrum}). None of these clouds show signs of external
or internal heating, for instance by an embedded star.

\par Peak line intensities are modest and line profiles are narrow
(line widths 0.7 to 1.3 $\mathrm{km\,s^{-1}}$), which is characteristic of
relatively cold, quiescent molecular gas. The $J$=1-0
$^{12}$CO/$^{13}$CO isotopolog ratios $R_{10}$ range from
$4.5\pm0.5$ to $7.0\pm0.8$. The small ``blue'' cloud at the center of
Simeis~57 has a negative velocity $V_{LSR}$ = $-10.9$ km s$^{-1}$
(cf. Fig.\,\ref{fig:blt_co_maps}), which is the same as
the recombination line velocity of the ionized gas $V_{LSR}$ =
$-12\, \mathrm{km\,s^{-1}}$ \citep{Pipenbrink1988}. Its isotopolog
ratio is higher ($R_{10}$ = $11.0\pm0.3$), which suggests a lower CO
optical depth.  

% Figure 11 Extinction map

\begin{figure*}
\includegraphics[width=1\textwidth]{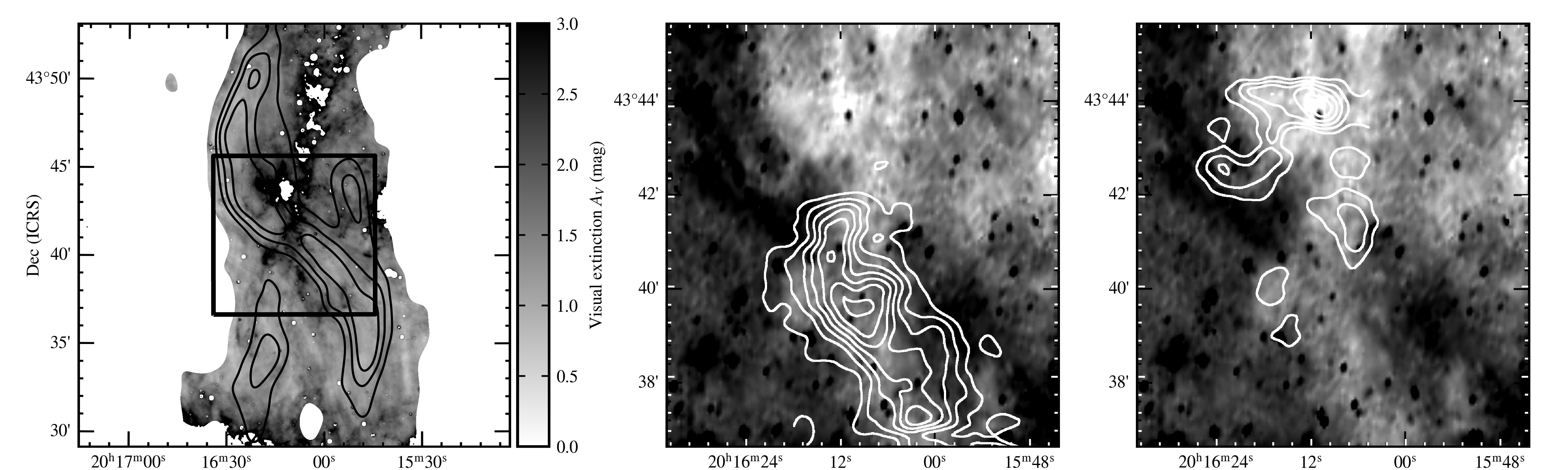}
\centering
\caption{Visual extinction map and millimeter CO data of Simeis 57.   Left: Extinction expressed in visual magnitudes across
  Simeis~57. Radio continuum contours
  $S_{1420} = 20,25,30\, \mathrm{mJy \, arcmin^{-2}}$ are
  superposed. The box indicates the outline of the two panels to the
  right. Center: JCMT $^{13}$CO contours (+5 km/s component)
  superposed on the blue digital Palomar Sky Survey (DSS)
  image. Right: JCMT $^{13}$CO contours (-10 km/s component)
  superposed on blue DSS image.
  }
\label{fig:co_extinction_map}
\end{figure*}

\par The more detailed JCMT maps show
(Fig.\,\ref{fig:jcmt_co_maps})
the CO clouds coinciding closely with individual absorption
patches in the visual image
(Fig.\,\ref{fig:co_extinction_map}). Especially, the blue cloudlets
at the same radial velocity as the optical nebula have clear
counterparts in well-defined extinction patches. The elongated red
cloud coincides with the extinction feature extending from the nebular
center to the south along DWB~119. This cloud is also in front of the
nebula, but given its different velocity, it may be at an intervening
distance.

\par The multiple-transition CO data allow for the derivation of the
physical conditions of the molecular gas with the radiative transfer
code RADEX \citep{vanderTak2007}.  We assume an intrinsic isotopolog
ratio [$^{12}$CO]/[$^{13}$CO] of 60, a carbon abundance [C]/[H] = 8.45
\citep[cf.][]{arellanocrdova2020galactic}, and a carbon gas-phase
depletion factor of a third. The blue cloud core then represents a
relatively large column
($N_{H_2}\,=\,2\times10^{21}\,\mathrm{m^{-2}}$) of rather warm
($T_{kin}\,\geq\,30$ K) molecular gas with a low density
  ($n_{H_2}$ = 100 cm$^{-3}$) that is close to the nebular gas
density (Sect. 3.2). The dust and molecular gas of this cloud appears
to be integral part of the nebula. In contrast, the more extended
red cloud core has a higher density ($1000\,\mathrm{cm^{-3}}$) but a
lower column density ($\sim 1\times10^{21}\, \mathrm{cm^{-2}}$) with
a temperature unconstrained by the observations. With standard
gas-to-extinction ratios \citep{zhu2017_gastoextinction} the
blue cloud has an $A_{V}\,\approx\,2^{m}$ and the red cloud
$A_{V}\,\approx\,1^{m}$.  Most of the molecular gas and dust traced
by the CO clouds is in front of the nebula.
There are no signs of embedded sources of excitation and the molecular
gas column densities imply extinctions much too low for the
obscuration of luminous stars. Thus, the Simeis~57 complex is not
hiding its source of excitation and we must look for it in the
much larger surrounding field of fragmentary molecular cloudlets,
nebular filaments, and scattered stars.

\subsection{Stars in the Simeis~57 field}
\label{section_starsinfield}
\par The ultraviolet and optical images in
Figs.\,\ref{fig:image_spectrum} and \ref{fig:co_extinction_map} 
might suggest that Simeis~57 is associated with a small group of
luminous blue stars. It should be noted that several of the
stars seen in the UV
images can be recognized at longer wavelengths up to $4.6\mu$m in the
near-infrared image and some can even be seen in the mid-infrared at
$12\mu$m and $22\mu$m. Because the wavelength-dependent interstellar
extinction increases steeply toward the ultraviolet, the stars in the
UV images are unlikely to suffer much extinction either locally or in
the foreground.

\par Appendix A shows the continuum spectra and associated data of
20 prominent stars near Simeis~57 (13 of which can be
identified in Fig.\,\ref{fig:image_spectrum}). As argued there, most
of these stars are ruled out as a source of excitation for Simeis~57
on grounds of spectral type, brightness, or distance. No exciting star
can be identified close to the nebula. Only the star HD~193793 =
WR~140 also known as the variable star V1687~Cyg or the spectroscopic
binary SBC9 1232 \citep{Pourbaix2004} emerges as a suitable candidate,
albeit at the large projected distance $50'$ to the east of the nebula
(cf. Figs.\,\ref{fig:Ha_mosaic} and \ref{fig:bright_stars}). At the
stellar distance of $1.71\pm0.11$ kpc \citep{smith2012,rate2020}, 
this corresponds to a projected linear separation of
about 23 pc. If this star and Simeis~57 are associated, they are
roughly in the same plane on the sky because Simeis~57 is seen
projected close to the edge of the shell surrounding HD~193793 in the
radio and dust maps shown by \cite{Arnal2001}.

HD~193793 consists of an evolved O4-5 star ($M_V\,=\,-6.9^{m}$)
and a WC7p Wolf-Rayet star ($M_V\,=\,-6.3^{m}$) with respective
luminosities log $L(O5I)/\mathrm{L_{\odot}}$ = 6.20 and log
$L(WC7)/\mathrm{L_{\odot}}$ = 5.34 \citep{Williams2011}. It is
located in a void in the HI distribution \citep{Arnal2001} and
suffers a line-of-sight extinction $A_V\,=\,2.9$
\citep{Williams2011}. If Simeis~57 is at the same distance as
HD~193793, its gas depth as defined by the radio emission would be
$d\,=\,EM/n_e^2\,\sim\,2$ pc (uncertain by a factor of about two).

\par In that case it covers a solid angle of $\sim0.06$
steradian. This corresponds to blocking $0.5\%$ of the sky, also with
a factor of two uncertainty. We took the model Lyman continuum
photon fluxes of a WC7 star from \cite{Crowther_2007} and those of an
O5I star from \cite{Sternberg_2003} and scaled them with the
corresponding observed-to-model luminosity ratios. We find a combined
production rate log
$N_{Lyc}(star)\,=\, 50.12 \,\mathrm{ph \, s^{-1}}$. There
appears to be little or no absorbing gas and dust between the star and
the nebula \citep{Arnal2001}, so that an intercept rate
$N_{lyc}(neb)\,=\,10^{47.8}\,\mathrm{ph\, s^{-1}}$ is
predicted for the nebula.  This is practically identical to the rate
derived from the H$\alpha$ observations in Sect. 3.3. The intercepted
infrared luminosity should be about $L_{TIR}\,\sim\,10^{4}$
L$_{\odot}$, that is, about equal to the total luminosity of the nebula. If
Simeis~57 is either closer or more distant than WR140, the numbers
change somewhat but remain comfortably within the uncertainty
limits. A significant difference, for instance by an amount equal to
its projected distance, is unlikely because this would place the nebula
well beyond the limits of the shell marking the sphere of influence of
the star.

\par The binary HD 193793 = WR140 is thus the most likely source for
the excitation of the nebula.  In particular, the UV Lyman continuum
output of the WR star is consistent with the strong [OIII] emission in
the eastern half of the nebula and the expected output of the system is in
excellent agreement with the ionization rate implied by the
observations of the nebula.
%%%%END EDIT REFEREE
In several aspects, such as the evolved nature of the exciting star,
its outside location, the [OIII]/H$\alpha$ emission line asymmetry,
and the filamentary appearance, Simeis~57 much resembles the
California Nebula (NGC~1499) excited by the hot runaway star $\xi$
Persei (O7.5III).
An investigation into the dynamics of the nebular gas might
be rewarding in studies of the interstellar medium of the region
and the evolution of an early O/WC binary such as HD~193793.

\section{Conclusions}

In the previous paper on Simeis~57, \cite{israel2003} found that
the emission of the nebula is thermal, ruling out that it is a
supernova remnant. Its peculiar shape is suggestive of a rotating
outflow from either an evolving young object such as an HII region or an
old evolved object such as a planetary nebula. In either case, we would
expect to find a centrally located source of excitation.

This paper establishes that the nebula has a relatively low density on
the order of 100 cm$^{-3 }$ and suffers an extinction of not much
more than 2$^{m}$.  Optical spectroscopy reveals an excitation gradient
% decreasing from east to west,
with significant [OIII] emission east of the main body and very
little [OIII] emission to the west. The nebula is recognizable but not
prominent in mid-infrared and far-infrared images. There is no major
CO cloud complex but small (4-8 pc) CO clouds, all at
$V_{LSR}\,\approx\,+5$ km s$^{-1}$, are scattered across the
field. One of these clouds coincides with the nebula and another
fragmentary cloud at the nebular velocity of $V_{LSR}\,\approx\,-10$
km s$^{-1}$. They do not appear to be physically connected and both
are smaller than the ionized nebula. They have substantial
substructure. The CO data suggest molecular gas densities of 1000
cm$^{-3}$ and 100 cm$^{-3}$ and modest column densities $N(H_{2})$ =
1-2 $\times$ 10$^{21}$ cm$^{-2}$.  These data clearly establish that
Simeis~57 is not part of a larger star-forming complex, but is an isolated
object in a larger field filled with fragmentary gas and dust
clouds. No luminous stars are embedded in the dust nor are any hidden
by it; there are no central objects.

The larger field surrounding the nebula reveals only one probable
excitation source. This is the evolved binary HD 193793 consisting of
an O4-5 supergiant and a WC7 Wolf-Rayet star (WR~140) at a
well-established distance of $D = 1.71$ kpc and with a projected
distance of 23 pc to the nebula. Both its luminosity and the hardness
of its UV radiation appear sufficient to explain the excitation of
Simeis~57. Moreover, its location to the east of the nebula fits well
with the observed [OIII] emission asymmetry. It thus appears that the
nebula Simeis~57 consists of separate filaments and diffuse emission
that together only fortuitously produce the remarkably coherent
appearance of an outflow object.

\begin{acknowledgements}
We are indebted to Ardjan Sturm and Christian Groeneveld for their
help with the May 2017 INT-WFC observations and discussions. We also thank Fedde Fagginger Auer, Margot Leemker, Martijn Wilhelm and Tom Sweegers for conducting the 2016 INT-WFC observations.  

\par We thank the telescope operators (TOs) of the James Clerk Maxwell
Telescope (JCMT) for conducting the observations described in this
paper. At the time, the JCMT was operated by the Joint Astronomy
Centre (JAC) in Hilo, Hawaii, on behalf of the Particle Physics and
Astronomy Research Council (P-PARC) of the United Kingdom, the Netherlands
Organization for Scientific Research (NWO), and the National Research
Council (NRC) of Canada.

\par JB acknowledges support by Funda\c{c}\~ao para a Ci\^encia e a
Technologia (FCT) through the research grants UID/FIS/04434/2019,
UIDB/04434/2020, UIDP/04434/2020 and through the Investigador FCT
Contract No. IF/01654/2014/CP1215/CT0003.

\par This work has made use of data from the European Space Agency (ESA) mission
{\it Gaia} (\url{https://www.cosmos.esa.int/gaia}), processed by the {\it Gaia}
Data Processing and Analysis Consortium (DPAC,
\url{https://www.cosmos.esa.int/web/gaia/dpac/consortium}). Funding for the DPAC
has been provided by national institutions, in particular the institutions
participating in the {\it Gaia} Multilateral Agreement.

\par This research made use of Montage, funded by the National
Aeronautics and Space Administration's Earth Science Technology
Office, Computational Technologies Project, under Cooperative
Agreement Number NCC5-626 between NASA and the California Institute of
Technology. The code is maintained by the NASA/IPAC Infrared Science
Archive.

\par This research has made use of the SIMBAD database, operated at
CDS, Strasbourg, France

\end{acknowledgements}

\bibliographystyle{aa}
\bibliography{pln}

\begin{thebibliography}{70}
\expandafter\ifx\csname natexlab\endcsname\relax\def\natexlab#1{#1}\fi

\bibitem[{{Abrahamyan} {et~al.}(2015){Abrahamyan}, {Mickaelian}, \&
  {Knyazyan}}]{Abrahamyan2015A}
{Abrahamyan}, H.~V., {Mickaelian}, A.~M., \& {Knyazyan}, A.~V. 2015, Astronomy
  and Computing, 10, 99

\bibitem[{{Aguado} {et~al.}(2019){Aguado}, {Ahumada}, {Almeida}, {Anderson},
  {Andrews}, {Anguiano}, {Aquino Ort{\'\i}z}, {Arag{\'o}n-Salamanca},
  {Argudo-Fern{\'a}ndez}, {Aubert}, {Avila-Reese}, {Badenes}, {Barboza
  Rembold}, {Barger}, {Barrera-Ballesteros}, {Bates}, {Bautista}, {Beaton},
  {Beers}, {Belfiore}, {Bernardi}, {Bershady}, {Beutler}, {Bird}, {Bizyaev},
  {Blanc}, {Blanton}, {Blomqvist}, {Bolton}, {Boquien}, {Borissova}, {Bovy},
  {Brand t}, {Brinkmann}, {Brownstein}, {Bundy}, {Burgasser}, {Byler}, {Cano
  Diaz}, {Cappellari}, {Carrera}, {Cervantes Sodi}, {Chen}, {Cherinka}, {Choi},
  {Chung}, {Coffey}, {Comerford}, {Comparat}, {Covey}, {da Silva Ilha}, {da
  Costa}, {Dai}, {Damke}, {Darling}, {Davies}, {Dawson}, {de Sainte Agathe},
  {Deconto Machado}, {Del Moro}, {De Lee}, {Diamond-Stanic}, {Dom{\'\i}nguez
  S{\'a}nchez}, {Donor}, {Drory}, {du Mas des Bourboux}, {Duckworth}, {Dwelly},
  {Ebelke}, {Emsellem}, {Escoffier}, {Fern{\'a}ndez-Trincado}, {Feuillet},
  {Fischer}, {Fleming}, {Fraser-McKelvie}, {Freischlad}, {Frinchaboy}, {Fu},
  {Galbany}, {Garcia-Dias}, {Garc{\'\i}a-Hern{\'a}ndez}, {Garma Oehmichen},
  {Geimba Maia}, {Gil-Mar{\'\i}n}, {Grabowski}, {Gu}, {Guo}, {Ha},
  {Harrington}, {Hasselquist}, {Hayes}, {Hearty}, {Hernandez Toledo}, {Hicks},
  {Hogg}, {Holley-Bockelmann}, {Holtzman}, {Hsieh}, {Hunt}, {Hwang},
  {Ibarra-Medel}, {Jimenez Angel}, {Johnson}, {Jones}, {J{\"o}nsson},
  {Kinemuchi}, {Kollmeier}, {Krawczyk}, {Kreckel}, {Kruk}, {Lacerna}, {Lan},
  {Lane}, {Law}, {Lee}, {Li}, {Lian}, {Lin}, {Lin}, {Lintott}, {Long},
  {Longa-Pe{\~n}a}, {Mackereth}, {de la Macorra}, {Majewski}, {Malanushenko},
  {Manchado}, {Maraston}, {Mariappan}, {Marinelli}, {Marques-Chaves},
  {Masseron}, {Masters}, {McDermid}, {Medina Pe{\~n}a}, {Meneses-Goytia},
  {Merloni}, {Merrifield}, {Meszaros}, {Minniti}, {Minsley}, {Muna}, {Myers},
  {Nair}, {Correa do Nascimento}, {Newman}, {Nitschelm}, {Olmstead}, {Oravetz},
  {Oravetz}, {Ortega Minakata}, {Pace}, {Padilla}, {Palicio}, {Pan}, {Pan},
  {Parikh}, {Parker}, {Peirani}, {Penny}, {Percival}, {Perez-Fournon},
  {Peterken}, {Pinsonneault}, {Prakash}, {Raddick}, {Raichoor}, {Riffel},
  {Riffel}, {Rix}, {Robin}, {Roman-Lopes}, {Rose}, {Ross}, {Rossi}, {Rowlands},
  {Rubin}, {S{\'a}nchez}, {S{\'a}nchez-Gallego}, {Sayres}, {Schaefer},
  {Schiavon}, {Schimoia}, {Schlafly}, {Schlegel}, {Schneider}, {Schultheis},
  {Seo}, {Shamsi}, {Shao}, {Shen}, {Shetty}, {Simonian}, {Smethurst}, {Sobeck},
  {Souter}, {Spindler}, {Stark}, {Stassun}, {Steinmetz}, {Storchi-Bergmann},
  {Stringfellow}, {Su{\'a}rez}, {Sun}, {Taghizadeh-Popp}, {Talbot}, {Tayar},
  {Thakar}, {Thomas}, {Tissera}, {Tojeiro}, {Troup}, {Unda-Sanzana},
  {Valenzuela}, {Vargas-Maga{\~n}a}, {V{\'a}zquez-Mata}, {Wake}, {Weaver},
  {Weijmans}, {Westfall}, {Wild}, {Wilson}, {Woods}, {Yan}, {Yang}, {Zamora},
  {Zasowski}, {Zhang}, {Zheng}, {Zheng}, {Zhu}, {Zinn}, \& {Zou}}]{Aguado2019}
{Aguado}, D.~S., {Ahumada}, R., {Almeida}, A., {et~al.} 2019, \apjs, 240, 23

\bibitem[{Arellano-Córdova {et~al.}(2020)Arellano-Córdova, Esteban,
  García-Rojas, \& Méndez-Delgado}]{arellanocrdova2020galactic}
Arellano-Córdova, K.~Z., Esteban, C., García-Rojas, J., \& Méndez-Delgado,
  J.~E. 2020, The Galactic radial abundance gradients of C, N, O, Ne, S, Cl and
  Ar from deep spectra of H II regions

\bibitem[{Armengot {et~al.}(2014)Armengot, S{\'a}nchez, L{\'o}pez-Santiago, \&
  de~Castro}]{armengot2014mosaix}
Armengot, M., S{\'a}nchez, N., L{\'o}pez-Santiago, J., \& de~Castro, A. I.~G.
  2014, Astrophysics and Space Science, 354, 113

\bibitem[{{Arnal}(2001)}]{Arnal2001}
{Arnal}, E.~M. 2001, \aj, 121, 413

\bibitem[{{Aumann} {et~al.}(1990){Aumann}, {Fowler}, \& {Melnyk}}]{Aumann1990}
{Aumann}, H.~H., {Fowler}, J.~W., \& {Melnyk}, M. 1990, \aj, 99, 1674

\bibitem[{{Bally} {et~al.}(1987){Bally}, {Langer}, {Stark}, \&
  {Wilson}}]{Bally1987}
{Bally}, J., {Langer}, W.~D., {Stark}, A.~A., \& {Wilson}, R.~W. 1987, \apjl,
  312, L45

\bibitem[{{Barentsen} {et~al.}(2014){Barentsen}, {Farnhill}, {Drew},
  {Gonzalez-Solares}, {Greimel}, {Irwin}, {Miszalski}, {Ruhland}, {Groot},
  {Mampaso}, {Sale}, {Henden}, {Aungwerojwit}, {Barlow}, {Carter}, {Corradi},
  {Drake}, {Eisloffel}, {Fabregat}, {Gansicke}, {Gentile Fusillo}, {Greiss},
  {Hales}, {Hodgkin}, {Huckvale}, {Irwin}, {King}, {Knigge}, {Kupfer},
  {Lagadec}, {Lennon}, {Lewis}, {Mohr-Smith}, {Morris}, {Naylor}, {Parker},
  {Phillipps}, {Pyrzas}, {Raddi}, {Roelofs}, {Rodriguez-Gil}, {Sabin},
  {Scaringi}, {Steeghs}, {Suso}, {Tata}, {Unruh}, {van Roestel}, {Viironen},
  {Vink}, {Walton}, {Wright}, \& {Zijlstra}}]{Barentsen2014}
{Barentsen}, G., {Farnhill}, H.~J., {Drew}, J.~E., {et~al.} 2014, VizieR Online
  Data Catalog, II/321

\bibitem[{Battersby {et~al.}(2011)Battersby, Bally, Ginsburg, Bernard, Brunt,
  Fuller, Martin, Molinari, Mottram, Peretto,
  {et~al.}}]{battersby2011characterizing}
Battersby, C., Bally, J., Ginsburg, A., {et~al.} 2011, Astronomy \&
  Astrophysics, 535, A128

\bibitem[{{Bertin}(2006)}]{2006ASPC..351..112B}
{Bertin}, E. 2006, in Astronomical Society of the Pacific Conference Series,
  Vol. 351, Astronomical Data Analysis Software and Systems XV, ed.
  C.~{Gabriel}, C.~{Arviset}, D.~{Ponz}, \& S.~{Enrique}, 112

\bibitem[{{Bertin} \& {Arnouts}(1996)}]{1996A&AS..117..393B}
{Bertin}, E. \& {Arnouts}, S. 1996, \aaps, 117, 393

\bibitem[{{Cardelli} {et~al.}(1989){Cardelli}, {Clayton}, \&
  {Mathis}}]{Cardelli1989}
{Cardelli}, J.~A., {Clayton}, G.~C., \& {Mathis}, J.~S. 1989, \apj, 345, 245

\bibitem[{{Cash} {et~al.}(1980){Cash}, {Charles}, {Bowyer}, {Walter},
  {Garmire}, \& {Riegler}}]{cash1980}
{Cash}, W., {Charles}, P., {Bowyer}, S., {et~al.} 1980, \apjl, 238, L71

\bibitem[{{Crowther}(2005)}]{Crowther2005}
{Crowther}, P.~A. 2005, in Massive Star Birth: A Crossroads of Astrophysics,
  ed. R.~{Cesaroni}, M.~{Felli}, E.~{Churchwell}, \& M.~{Walmsley}, Vol. 227,
  389--396

\bibitem[{Crowther(2007)}]{Crowther_2007}
Crowther, P.~A. 2007, Annual Review of Astronomy and Astrophysics, 45,
  177–219

\bibitem[{{Cutri} {et~al.}(2003){Cutri}, {Skrutskie}, {van Dyk}, {Beichman},
  {Carpenter}, {Chester}, {Cambresy}, {Evans}, {Fowler}, {Gizis}, {Howard},
  {Huchra}, {Jarrett}, {Kopan}, {Kirkpatrick}, {Light}, {Marsh}, {McCallon},
  {Schneider}, {Stiening}, {Sykes}, {Weinberg}, {Wheaton}, {Wheelock}, \&
  {Zacarias}}]{Cutri2003}
{Cutri}, R.~M., {Skrutskie}, M.~F., {van Dyk}, S., {et~al.} 2003, VizieR Online
  Data Catalog, II/246

\bibitem[{{Doi} {et~al.}(2015){Doi}, {Takita}, {Ootsubo}, {Arimatsu}, {Tanaka},
  {Kitamura}, {Kawada}, {Matsuura}, {Nakagawa}, {Morishima}, {Hattori},
  {Komugi}, {White}, {Ikeda}, {Kato}, {Chinone}, {Etxaluze}, \&
  {Cypriano}}]{AkariPaper}
{Doi}, Y., {Takita}, S., {Ootsubo}, T., {et~al.} 2015, \pasj, 67, 50

\bibitem[{{Draine}(2011)}]{2011piim.book.....D}
{Draine}, B.~T. 2011, {Physics of the Interstellar and Intergalactic Medium}
  (Princeton University Press)

\bibitem[{{Drew} {et~al.}(2005){Drew}, {Greimel}, {Irwin}, {Aungwerojwit},
  {Barlow}, {Corradi}, {Drake}, {G{\"a}nsicke}, {Groot}, {Hales}, {Hopewell},
  {Irwin}, {Knigge}, {Leisy}, {Lennon}, {Mampaso}, {Masheder}, {Matsuura},
  {Morales-Rueda}, {Morris}, {Parker}, {Phillipps}, {Rodriguez-Gil}, {Roelofs},
  {Skillen}, {Sokoloski}, {Steeghs}, {Unruh}, {Viironen}, {Vink}, {Walton},
  {Witham}, {Wright}, {Zijlstra}, \& {Zurita}}]{iphas1}
{Drew}, J.~E., {Greimel}, R., {Irwin}, M.~J., {et~al.} 2005, \mnras, 362, 753

\bibitem[{{Egan} {et~al.}(2003){Egan}, {Price}, {Kraemer}, {Mizuno}, {Carey},
  {Wright}, {Engelke}, {Cohen}, \& {Gugliotti}}]{Egan2003}
{Egan}, M.~P., {Price}, S.~D., {Kraemer}, K.~E., {et~al.} 2003, VizieR Online
  Data Catalog, V/114

\bibitem[{{Egret} {et~al.}(1992){Egret}, {Didelon}, {McLean}, {Russell}, \&
  {Turon}}]{Egret1992A}
{Egret}, D., {Didelon}, P., {McLean}, B.~J., {Russell}, J.~L., \& {Turon}, C.
  1992, \aap, 258, 217

\bibitem[{Erben {et~al.}(2005)Erben, Schirmer, Dietrich, Cordes, Haberzettl,
  Hetterscheidt, Hildebrandt, Schmithuesen, Schneider, Simon, Deul, Hook,
  Kaiser, Radovich, Benoist, Nonino, Olsen, Prandoni, Wichmann, Zaggia, Bomans,
  Dettmar, \& Miralles}]{ErbenTheli}
Erben, T., Schirmer, M., Dietrich, J.~P., {et~al.} 2005, Astronomische
  Nachrichten, 326, 432

\bibitem[{{Gaia Collaboration} {et~al.}(2018{\natexlab{a}}){Gaia
  Collaboration}, {Brown}, {Vallenari}, {Prusti}, {de Bruijne}, {Babusiaux},
  {Bailer-Jones}, {Biermann}, {Evans}, {Eyer}, {Jansen}, {Jordi}, {Klioner},
  {Lammers}, {Lindegren}, {Luri}, {Mignard}, {Panem}, {Pourbaix}, {Randich},
  {Sartoretti}, {Siddiqui}, {Soubiran}, {van Leeuwen}, {Walton}, {Arenou},
  {Bastian}, {Cropper}, {Drimmel}, {Katz}, {Lattanzi}, {Bakker}, {Cacciari},
  {Casta{\~n}eda}, {Chaoul}, {Cheek}, {De Angeli}, {Fabricius}, {Guerra},
  {Holl}, {Masana}, {Messineo}, {Mowlavi}, {Nienartowicz}, {Panuzzo},
  {Portell}, {Riello}, {Seabroke}, {Tanga}, {Th{\'e}venin}, {Gracia-Abril},
  {Comoretto}, {Garcia-Reinaldos}, {Teyssier}, {Altmann}, {Andrae}, {Audard},
  {Bellas-Velidis}, {Benson}, {Berthier}, {Blomme}, {Burgess}, {Busso},
  {Carry}, {Cellino}, {Clementini}, {Clotet}, {Creevey}, {Davidson}, {De
  Ridder}, {Delchambre}, {Dell'Oro}, {Ducourant},
  {Fern{\'a}ndez-Hern{\'a}ndez}, {Fouesneau}, {Fr{\'e}mat}, {Galluccio},
  {Garc{\'\i}a-Torres}, {Gonz{\'a}lez-N{\'u}{\~n}ez}, {Gonz{\'a}lez-Vidal},
  {Gosset}, {Guy}, {Halbwachs}, {Hambly}, {Harrison}, {Hern{\'a}ndez},
  {Hestroffer}, {Hodgkin}, {Hutton}, {Jasniewicz}, {Jean-Antoine-Piccolo},
  {Jordan}, {Korn}, {Krone-Martins}, {Lanzafame}, {Lebzelter}, {L{\"o}ffler},
  {Manteiga}, {Marrese}, {Mart{\'\i}n-Fleitas}, {Moitinho}, {Mora}, {Muinonen},
  {Osinde}, {Pancino}, {Pauwels}, {Petit}, {Recio-Blanco}, {Richards},
  {Rimoldini}, {Robin}, {Sarro}, {Siopis}, {Smith}, {Sozzetti}, {S{\"u}veges},
  {Torra}, {van Reeven}, {Abbas}, {Abreu Aramburu}, {Accart}, {Aerts},
  {Altavilla}, {{\'A}lvarez}, {Alvarez}, {Alves}, {Anderson}, {Andrei},
  {Anglada Varela}, {Antiche}, {Antoja}, {Arcay}, {Astraatmadja}, {Bach},
  {Baker}, {Balaguer-N{\'u}{\~n}ez}, {Balm}, {Barache}, {Barata}, {Barbato},
  {Barblan}, {Barklem}, {Barrado}, {Barros}, {Barstow}, {Bartholom{\'e}
  Mu{\~n}oz}, {Bassilana}, {Becciani}, {Bellazzini}, {Berihuete}, {Bertone},
  {Bianchi}, {Bienaym{\'e}}, {Blanco-Cuaresma}, {Boch}, {Boeche}, {Bombrun},
  {Borrachero}, {Bossini}, {Bouquillon}, {Bourda}, {Bragaglia}, {Bramante},
  {Breddels}, {Bressan}, {Brouillet}, {Br{\"u}semeister}, {Brugaletta},
  {Bucciarelli}, {Burlacu}, {Busonero}, {Butkevich}, {Buzzi}, {Caffau},
  {Cancelliere}, {Cannizzaro}, {Cantat-Gaudin}, {Carballo}, {Carlucci},
  {Carrasco}, {Casamiquela}, {Castellani}, {Castro-Ginard}, {Charlot},
  {Chemin}, {Chiavassa}, {Cocozza}, {Costigan}, {Cowell}, {Crifo}, {Crosta},
  {Crowley}, {Cuypers}, {Dafonte}, {Damerdji}, {Dapergolas}, {David}, {David},
  {de Laverny}, {De Luise}, {De March}, {de Martino}, {de Souza}, {de Torres},
  {Debosscher}, {del Pozo}, {Delbo}, {Delgado}, {Delgado}, {Di Matteo},
  {Diakite}, {Diener}, {Distefano}, {Dolding}, {Drazinos}, {Dur{\'a}n},
  {Edvardsson}, {Enke}, {Eriksson}, {Esquej}, {Eynard Bontemps}, {Fabre},
  {Fabrizio}, {Faigler}, {Falc{\~a}o}, {Farr{\`a}s Casas}, {Federici},
  {Fedorets}, {Fernique}, {Figueras}, {Filippi}, {Findeisen}, {Fonti},
  {Fraile}, {Fraser}, {Fr{\'e}zouls}, {Gai}, {Galleti}, {Garabato},
  {Garc{\'\i}a-Sedano}, {Garofalo}, {Garralda}, {Gavel}, {Gavras}, {Gerssen},
  {Geyer}, {Giacobbe}, {Gilmore}, {Girona}, {Giuffrida}, {Glass}, {Gomes},
  {Granvik}, {Gueguen}, {Guerrier}, {Guiraud}, {Guti{\'e}rrez-S{\'a}nchez},
  {Haigron}, {Hatzidimitriou}, {Hauser}, {Haywood}, {Heiter}, {Helmi}, {Heu},
  {Hilger}, {Hobbs}, {Hofmann}, {Holland}, {Huckle}, {Hypki}, {Icardi},
  {Jan{\ss}en}, {Jevardat de Fombelle}, {Jonker}, {Juh{\'a}sz}, {Julbe},
  {Karampelas}, {Kewley}, {Klar}, {Kochoska}, {Kohley}, {Kolenberg},
  {Kontizas}, {Kontizas}, {Koposov}, {Kordopatis}, {Kostrzewa-Rutkowska},
  {Koubsky}, {Lambert}, {Lanza}, {Lasne}, {Lavigne}, {Le Fustec}, {Le
  Poncin-Lafitte}, {Lebreton}, {Leccia}, {Leclerc}, {Lecoeur-Taibi},
  {Lenhardt}, {Leroux}, {Liao}, {Licata}, {Lindstr{\o}m}, {Lister}, {Livanou},
  {Lobel}, {L{\'o}pez}, {Managau}, {Mann}, {Mantelet}, {Marchal}, {Marchant},
  {Marconi}, {Marinoni}, {Marschalk{\'o}}, {Marshall}, {Martino}, {Marton},
  {Mary}, {Massari}, {Matijevi{\v{c}}}, {Mazeh}, {McMillan}, {Messina},
  {Michalik}, {Millar}, {Molina}, {Molinaro}, {Moln{\'a}r}, {Montegriffo},
  {Mor}, {Morbidelli}, {Morel}, {Morris}, {Mulone}, {Muraveva}, {Musella},
  {Nelemans}, {Nicastro}, {Noval}, {O'Mullane}, {Ord{\'e}novic},
  {Ord{\'o}{\~n}ez-Blanco}, {Osborne}, {Pagani}, {Pagano}, {Pailler},
  {Palacin}, {Palaversa}, {Panahi}, {Pawlak}, {Piersimoni}, {Pineau}, {Plachy},
  {Plum}, {Poggio}, {Poujoulet}, {Pr{\v{s}}a}, {Pulone}, {Racero}, {Ragaini},
  {Rambaux}, {Ramos-Lerate}, {Regibo}, {Reyl{\'e}}, {Riclet}, {Ripepi}, {Riva},
  {Rivard}, {Rixon}, {Roegiers}, {Roelens}, {Romero-G{\'o}mez}, {Rowell},
  {Royer}, {Ruiz-Dern}, {Sadowski}, {Sagrist{\`a} Sell{\'e}s}, {Sahlmann},
  {Salgado}, {Salguero}, {Sanna}, {Santana-Ros}, {Sarasso}, {Savietto},
  {Schultheis}, {Sciacca}, {Segol}, {Segovia}, {S{\'e}gransan}, {Shih},
  {Siltala}, {Silva}, {Smart}, {Smith}, {Solano}, {Solitro}, {Sordo}, {Soria
  Nieto}, {Souchay}, {Spagna}, {Spoto}, {Stampa}, {Steele},
  {Steidelm{\"u}ller}, {Stephenson}, {Stoev}, {Suess}, {Surdej}, {Szabados},
  {Szegedi-Elek}, {Tapiador}, {Taris}, {Tauran}, {Taylor}, {Teixeira},
  {Terrett}, {Teyssand ier}, {Thuillot}, {Titarenko}, {Torra Clotet}, {Turon},
  {Ulla}, {Utrilla}, {Uzzi}, {Vaillant}, {Valentini}, {Valette}, {van Elteren},
  {Van Hemelryck}, {van Leeuwen}, {Vaschetto}, {Vecchiato}, {Veljanoski},
  {Viala}, {Vicente}, {Vogt}, {von Essen}, {Voss}, {Votruba}, {Voutsinas},
  {Walmsley}, {Weiler}, {Wertz}, {Wevers}, {Wyrzykowski}, {Yoldas},
  {{\v{Z}}erjal}, {Ziaeepour}, {Zorec}, {Zschocke}, {Zucker}, {Zurbach}, \&
  {Zwitter}}]{gaia_data}
{Gaia Collaboration}, {Brown}, A.~G.~A., {Vallenari}, A., {et~al.}
  2018{\natexlab{a}}, \aap, 616, A1

\bibitem[{{Gaia Collaboration} {et~al.}(2018{\natexlab{b}}){Gaia
  Collaboration}, {Brown, A. G. A.}, {Vallenari, A.}, {Prusti, T.}, {de
  Bruijne, J. H. J.}, {Babusiaux, C.}, {Bailer-Jones, C. A. L.}, {Biermann,
  M.}, {Evans, D. W.}, {Eyer, L.}, {Jansen, F.}, {Jordi, C.}, {Klioner, S. A.},
  {Lammers, U.}, {Lindegren, L.}, {Luri, X.}, {Mignard, F.}, {Panem, C.},
  {Pourbaix, D.}, {Randich, S.}, {Sartoretti, P.}, {Siddiqui, H. I.},
  {Soubiran, C.}, {van Leeuwen, F.}, {Walton, N. A.}, {Arenou, F.}, {Bastian,
  U.}, {Cropper, M.}, {Drimmel, R.}, {Katz, D.}, {Lattanzi, M. G.}, {Bakker,
  J.}, {Cacciari, C.}, {Casta\~neda, J.}, {Chaoul, L.}, {Cheek, N.}, {De
  Angeli, F.}, {Fabricius, C.}, {Guerra, R.}, {Holl, B.}, {Masana, E.},
  {Messineo, R.}, {Mowlavi, N.}, {Nienartowicz, K.}, {Panuzzo, P.}, {Portell,
  J.}, {Riello, M.}, {Seabroke, G. M.}, {Tanga, P.}, {Th\'evenin, F.},
  {Gracia-Abril, G.}, {Comoretto, G.}, {Garcia-Reinaldos, M.}, {Teyssier, D.},
  {Altmann, M.}, {Andrae, R.}, {Audard, M.}, {Bellas-Velidis, I.}, {Benson,
  K.}, {Berthier, J.}, {Blomme, R.}, {Burgess, P.}, {Busso, G.}, {Carry, B.},
  {Cellino, A.}, {Clementini, G.}, {Clotet, M.}, {Creevey, O.}, {Davidson, M.},
  {De Ridder, J.}, {Delchambre, L.}, {Dell\'{}Oro, A.}, {Ducourant, C.},
  {Fern\'andez-Hern\'andez, J.}, {Fouesneau, M.}, {Fr\'emat, Y.}, {Galluccio,
  L.}, {Garc\'{\i}a-Torres, M.}, {Gonz\'alez-N\'u\~nez, J.}, {Gonz\'alez-Vidal,
  J. J.}, {Gosset, E.}, {Guy, L. P.}, {Halbwachs, J.-L.}, {Hambly, N. C.},
  {Harrison, D. L.}, {Hern\'andez, J.}, {Hestroffer, D.}, {Hodgkin, S. T.},
  {Hutton, A.}, {Jasniewicz, G.}, {Jean-Antoine-Piccolo, A.}, {Jordan, S.},
  {Korn, A. J.}, {Krone-Martins, A.}, {Lanzafame, A. C.}, {Lebzelter, T.},
  {L\"offler, W.}, {Manteiga, M.}, {Marrese, P. M.}, {Mart\'{\i}n-Fleitas, J.
  M.}, {Moitinho, A.}, {Mora, A.}, {Muinonen, K.}, {Osinde, J.}, {Pancino, E.},
  {Pauwels, T.}, {Petit, J.-M.}, {Recio-Blanco, A.}, {Richards, P. J.},
  {Rimoldini, L.}, {Robin, A. C.}, {Sarro, L. M.}, {Siopis, C.}, {Smith, M.},
  {Sozzetti, A.}, {S\"uveges, M.}, {Torra, J.}, {van Reeven, W.}, {Abbas, U.},
  {Abreu Aramburu, A.}, {Accart, S.}, {Aerts, C.}, {Altavilla, G.}, {\'Alvarez,
  M. A.}, {Alvarez, R.}, {Alves, J.}, {Anderson, R. I.}, {Andrei, A. H.},
  {Anglada Varela, E.}, {Antiche, E.}, {Antoja, T.}, {Arcay, B.},
  {Astraatmadja, T. L.}, {Bach, N.}, {Baker, S. G.}, {Balaguer-N\'u\~nez, L.},
  {Balm, P.}, {Barache, C.}, {Barata, C.}, {Barbato, D.}, {Barblan, F.},
  {Barklem, P. S.}, {Barrado, D.}, {Barros, M.}, {Barstow, M. A.},
  {Bartholom\'e Mu\~noz, S.}, {Bassilana, J.-L.}, {Becciani, U.}, {Bellazzini,
  M.}, {Berihuete, A.}, {Bertone, S.}, {Bianchi, L.}, {Bienaym\'e, O.},
  {Blanco-Cuaresma, S.}, {Boch, T.}, {Boeche, C.}, {Bombrun, A.}, {Borrachero,
  R.}, {Bossini, D.}, {Bouquillon, S.}, {Bourda, G.}, {Bragaglia, A.},
  {Bramante, L.}, {Breddels, M. A.}, {Bressan, A.}, {Brouillet, N.},
  {Br\"usemeister, T.}, {Brugaletta, E.}, {Bucciarelli, B.}, {Burlacu, A.},
  {Busonero, D.}, {Butkevich, A. G.}, {Buzzi, R.}, {Caffau, E.}, {Cancelliere,
  R.}, {Cannizzaro, G.}, {Cantat-Gaudin, T.}, {Carballo, R.}, {Carlucci, T.},
  {Carrasco, J. M.}, {Casamiquela, L.}, {Castellani, M.}, {Castro-Ginard, A.},
  {Charlot, P.}, {Chemin, L.}, {Chiavassa, A.}, {Cocozza, G.}, {Costigan, G.},
  {Cowell, S.}, {Crifo, F.}, {Crosta, M.}, {Crowley, C.}, {Cuypers+, J.},
  {Dafonte, C.}, {Damerdji, Y.}, {Dapergolas, A.}, {David, P.}, {David, M.},
  {de Laverny, P.}, {De Luise, F.}, {De March, R.}, {de Martino, D.}, {de
  Souza, R.}, {de Torres, A.}, {Debosscher, J.}, {del Pozo, E.}, {Delbo, M.},
  {Delgado, A.}, {Delgado, H. E.}, {Di Matteo, P.}, {Diakite, S.}, {Diener,
  C.}, {Distefano, E.}, {Dolding, C.}, {Drazinos, P.}, {Dur\'an, J.},
  {Edvardsson, B.}, {Enke, H.}, {Eriksson, K.}, {Esquej, P.}, {Eynard Bontemps,
  G.}, {Fabre, C.}, {Fabrizio, M.}, {Faigler, S.}, {Falc\~ao, A. J.}, {Farr\`as
  Casas, M.}, {Federici, L.}, {Fedorets, G.}, {Fernique, P.}, {Figueras, F.},
  {Filippi, F.}, {Findeisen, K.}, {Fonti, A.}, {Fraile, E.}, {Fraser, M.},
  {Fr\'ezouls, B.}, {Gai, M.}, {Galleti, S.}, {Garabato, D.},
  {Garc\'{\i}a-Sedano, F.}, {Garofalo, A.}, {Garralda, N.}, {Gavel, A.},
  {Gavras, P.}, {Gerssen, J.}, {Geyer, R.}, {Giacobbe, P.}, {Gilmore, G.},
  {Girona, S.}, {Giuffrida, G.}, {Glass, F.}, {Gomes, M.}, {Granvik, M.},
  {Gueguen, A.}, {Guerrier, A.}, {Guiraud, J.}, {Guti\'errez-S\'anchez, R.},
  {Haigron, R.}, {Hatzidimitriou, D.}, {Hauser, M.}, {Haywood, M.}, {Heiter,
  U.}, {Helmi, A.}, {Heu, J.}, {Hilger, T.}, {Hobbs, D.}, {Hofmann, W.},
  {Holland, G.}, {Huckle, H. E.}, {Hypki, A.}, {Icardi, V.}, {Jan\ss{}en, K.},
  {Jevardat de Fombelle, G.}, {Jonker, P. G.}, {Juh\'asz, \'A. L.}, {Julbe,
  F.}, {Karampelas, A.}, {Kewley, A.}, {Klar, J.}, {Kochoska, A.}, {Kohley,
  R.}, {Kolenberg, K.}, {Kontizas, M.}, {Kontizas, E.}, {Koposov, S. E.},
  {Kordopatis, G.}, {Kostrzewa-Rutkowska, Z.}, {Koubsky, P.}, {Lambert, S.},
  {Lanza, A. F.}, {Lasne, Y.}, {Lavigne, J.-B.}, {Le Fustec, Y.}, {Le
  Poncin-Lafitte, C.}, {Lebreton, Y.}, {Leccia, S.}, {Leclerc, N.},
  {Lecoeur-Taibi, I.}, {Lenhardt, H.}, {Leroux, F.}, {Liao, S.}, {Licata, E.},
  {Lindstr\o{}m, H. E. P.}, {Lister, T. A.}, {Livanou, E.}, {Lobel, A.},
  {L\'opez, M.}, {Managau, S.}, {Mann, R. G.}, {Mantelet, G.}, {Marchal, O.},
  {Marchant, J. M.}, {Marconi, M.}, {Marinoni, S.}, {Marschalk\'o, G.},
  {Marshall, D. J.}, {Martino, M.}, {Marton, G.}, {Mary, N.}, {Massari, D.},
  {Matijevic, G.}, {Mazeh, T.}, {McMillan, P. J.}, {Messina, S.}, {Michalik,
  D.}, {Millar, N. R.}, {Molina, D.}, {Molinaro, R.}, {Moln\'ar, L.},
  {Montegriffo, P.}, {Mor, R.}, {Morbidelli, R.}, {Morel, T.}, {Morris, D.},
  {Mulone, A. F.}, {Muraveva, T.}, {Musella, I.}, {Nelemans, G.}, {Nicastro,
  L.}, {Noval, L.}, {O\'{}Mullane, W.}, {Ord\'enovic, C.}, {Ord\'o\~nez-Blanco,
  D.}, {Osborne, P.}, {Pagani, C.}, {Pagano, I.}, {Pailler, F.}, {Palacin, H.},
  {Palaversa, L.}, {Panahi, A.}, {Pawlak, M.}, {Piersimoni, A. M.}, {Pineau,
  F.-X.}, {Plachy, E.}, {Plum, G.}, {Poggio, E.}, {Poujoulet, E.}, {Prsa, A.},
  {Pulone, L.}, {Racero, E.}, {Ragaini, S.}, {Rambaux, N.}, {Ramos-Lerate, M.},
  {Regibo, S.}, {Reyl\'e, C.}, {Riclet, F.}, {Ripepi, V.}, {Riva, A.}, {Rivard,
  A.}, {Rixon, G.}, {Roegiers, T.}, {Roelens, M.}, {Romero-G\'omez, M.},
  {Rowell, N.}, {Royer, F.}, {Ruiz-Dern, L.}, {Sadowski, G.}, {Sagrist\`a
  Sell\'es, T.}, {Sahlmann, J.}, {Salgado, J.}, {Salguero, E.}, {Sanna, N.},
  {Santana-Ros, T.}, {Sarasso, M.}, {Savietto, H.}, {Schultheis, M.}, {Sciacca,
  E.}, {Segol, M.}, {Segovia, J. C.}, {S\'egransan, D.}, {Shih, I-C.},
  {Siltala, L.}, {Silva, A. F.}, {Smart, R. L.}, {Smith, K. W.}, {Solano, E.},
  {Solitro, F.}, {Sordo, R.}, {Soria Nieto, S.}, {Souchay, J.}, {Spagna, A.},
  {Spoto, F.}, {Stampa, U.}, {Steele, I. A.}, {Steidelm\"uller, H.},
  {Stephenson, C. A.}, {Stoev, H.}, {Suess, F. F.}, {Surdej, J.}, {Szabados,
  L.}, {Szegedi-Elek, E.}, {Tapiador, D.}, {Taris, F.}, {Tauran, G.}, {Taylor,
  M. B.}, {Teixeira, R.}, {Terrett, D.}, {Teyssandier, P.}, {Thuillot, W.},
  {Titarenko, A.}, {Torra Clotet, F.}, {Turon, C.}, {Ulla, A.}, {Utrilla, E.},
  {Uzzi, S.}, {Vaillant, M.}, {Valentini, G.}, {Valette, V.}, {van Elteren,
  A.}, {Van Hemelryck, E.}, {van Leeuwen, M.}, {Vaschetto, M.}, {Vecchiato,
  A.}, {Veljanoski, J.}, {Viala, Y.}, {Vicente, D.}, {Vogt, S.}, {von Essen,
  C.}, {Voss, H.}, {Votruba, V.}, {Voutsinas, S.}, {Walmsley, G.}, {Weiler,
  M.}, {Wertz, O.}, {Wevers, T.}, {Wyrzykowski, L.}, {Yoldas, A.}, {Zerjal,
  M.}, {Ziaeepour, H.}, {Zorec, J.}, {Zschocke, S.}, {Zucker, S.}, {Zurbach,
  C.}, \& {Zwitter, T.}}]{refId0}
{Gaia Collaboration}, {Brown, A. G. A.}, {Vallenari, A.}, {et~al.}
  2018{\natexlab{b}}, A\&A, 616, A1

\bibitem[{{Gaia Collaboration} {et~al.}(2016){Gaia Collaboration}, {Prusti},
  {de Bruijne}, {Brown}, {Vallenari}, {Babusiaux}, {Bailer-Jones}, {Bastian},
  {Biermann}, {Evans}, {Eyer}, {Jansen}, {Jordi}, {Klioner}, {Lammers},
  {Lindegren}, {Luri}, {Mignard}, {Milligan}, {Panem}, {Poinsignon},
  {Pourbaix}, {Randich}, {Sarri}, {Sartoretti}, {Siddiqui}, {Soubiran},
  {Valette}, {van Leeuwen}, {Walton}, {Aerts}, {Arenou}, {Cropper}, {Drimmel},
  {H{\o}g}, {Katz}, {Lattanzi}, {O'Mullane}, {Grebel}, {Holland}, {Huc},
  {Passot}, {Bramante}, {Cacciari}, {Casta{\~n}eda}, {Chaoul}, {Cheek}, {De
  Angeli}, {Fabricius}, {Guerra}, {Hern{\'a}ndez}, {Jean-Antoine-Piccolo},
  {Masana}, {Messineo}, {Mowlavi}, {Nienartowicz}, {Ord{\'o}{\~n}ez-Blanco},
  {Panuzzo}, {Portell}, {Richards}, {Riello}, {Seabroke}, {Tanga},
  {Th{\'e}venin}, {Torra}, {Els}, {Gracia-Abril}, {Comoretto},
  {Garcia-Reinaldos}, {Lock}, {Mercier}, {Altmann}, {Andrae}, {Astraatmadja},
  {Bellas-Velidis}, {Benson}, {Berthier}, {Blomme}, {Busso}, {Carry},
  {Cellino}, {Clementini}, {Cowell}, {Creevey}, {Cuypers}, {Davidson}, {De
  Ridder}, {de Torres}, {Delchambre}, {Dell'Oro}, {Ducourant}, {Fr{\'e}mat},
  {Garc{\'\i}a-Torres}, {Gosset}, {Halbwachs}, {Hambly}, {Harrison}, {Hauser},
  {Hestroffer}, {Hodgkin}, {Huckle}, {Hutton}, {Jasniewicz}, {Jordan},
  {Kontizas}, {Korn}, {Lanzafame}, {Manteiga}, {Moitinho}, {Muinonen},
  {Osinde}, {Pancino}, {Pauwels}, {Petit}, {Recio-Blanco}, {Robin}, {Sarro},
  {Siopis}, {Smith}, {Smith}, {Sozzetti}, {Thuillot}, {van Reeven}, {Viala},
  {Abbas}, {Abreu Aramburu}, {Accart}, {Aguado}, {Allan}, {Allasia},
  {Altavilla}, {{\'A}lvarez}, {Alves}, {Anderson}, {Andrei}, {Anglada Varela},
  {Antiche}, {Antoja}, {Ant{\'o}n}, {Arcay}, {Atzei}, {Ayache}, {Bach},
  {Baker}, {Balaguer-N{\'u}{\~n}ez}, {Barache}, {Barata}, {Barbier}, {Barblan},
  {Baroni}, {Barrado y Navascu{\'e}s}, {Barros}, {Barstow}, {Becciani},
  {Bellazzini}, {Bellei}, {Bello Garc{\'\i}a}, {Belokurov}, {Bendjoya},
  {Berihuete}, {Bianchi}, {Bienaym{\'e}}, {Billebaud}, {Blagorodnova},
  {Blanco-Cuaresma}, {Boch}, {Bombrun}, {Borrachero}, {Bouquillon}, {Bourda},
  {Bouy}, {Bragaglia}, {Breddels}, {Brouillet}, {Br{\"u}semeister},
  {Bucciarelli}, {Budnik}, {Burgess}, {Burgon}, {Burlacu}, {Busonero}, {Buzzi},
  {Caffau}, {Cambras}, {Campbell}, {Cancelliere}, {Cantat-Gaudin}, {Carlucci},
  {Carrasco}, {Castellani}, {Charlot}, {Charnas}, {Charvet}, {Chassat},
  {Chiavassa}, {Clotet}, {Cocozza}, {Collins}, {Collins}, {Costigan}, {Crifo},
  {Cross}, {Crosta}, {Crowley}, {Dafonte}, {Damerdji}, {Dapergolas}, {David},
  {David}, {De Cat}, {de Felice}, {de Laverny}, {De Luise}, {De March}, {de
  Martino}, {de Souza}, {Debosscher}, {del Pozo}, {Delbo}, {Delgado},
  {Delgado}, {di Marco}, {Di Matteo}, {Diakite}, {Distefano}, {Dolding}, {Dos
  Anjos}, {Drazinos}, {Dur{\'a}n}, {Dzigan}, {Ecale}, {Edvardsson}, {Enke},
  {Erdmann}, {Escolar}, {Espina}, {Evans}, {Eynard Bontemps}, {Fabre},
  {Fabrizio}, {Faigler}, {Falc{\~a}o}, {Farr{\`a}s Casas}, {Faye}, {Federici},
  {Fedorets}, {Fern{\'a}ndez-Hern{\'a}ndez}, {Fernique}, {Fienga}, {Figueras},
  {Filippi}, {Findeisen}, {Fonti}, {Fouesneau}, {Fraile}, {Fraser}, {Fuchs},
  {Furnell}, {Gai}, {Galleti}, {Galluccio}, {Garabato}, {Garc{\'\i}a-Sedano},
  {Gar{\'e}}, {Garofalo}, {Garralda}, {Gavras}, {Gerssen}, {Geyer}, {Gilmore},
  {Girona}, {Giuffrida}, {Gomes}, {Gonz{\'a}lez-Marcos},
  {Gonz{\'a}lez-N{\'u}{\~n}ez}, {Gonz{\'a}lez-Vidal}, {Granvik}, {Guerrier},
  {Guillout}, {Guiraud}, {G{\'u}rpide}, {Guti{\'e}rrez-S{\'a}nchez}, {Guy},
  {Haigron}, {Hatzidimitriou}, {Haywood}, {Heiter}, {Helmi}, {Hobbs},
  {Hofmann}, {Holl}, {Holland }, {Hunt}, {Hypki}, {Icardi}, {Irwin}, {Jevardat
  de Fombelle}, {Jofr{\'e}}, {Jonker}, {Jorissen}, {Julbe}, {Karampelas},
  {Kochoska}, {Kohley}, {Kolenberg}, {Kontizas}, {Koposov}, {Kordopatis},
  {Koubsky}, {Kowalczyk}, {Krone-Martins}, {Kudryashova}, {Kull}, {Bachchan},
  {Lacoste-Seris}, {Lanza}, {Lavigne}, {Le Poncin-Lafitte}, {Lebreton},
  {Lebzelter}, {Leccia}, {Leclerc}, {Lecoeur-Taibi}, {Lemaitre}, {Lenhardt},
  {Leroux}, {Liao}, {Licata}, {Lindstr{\o}m}, {Lister}, {Livanou}, {Lobel},
  {L{\"o}ffler}, {L{\'o}pez}, {Lopez-Lozano}, {Lorenz}, {Loureiro},
  {MacDonald}, {Magalh{\~a}es Fernandes}, {Managau}, {Mann}, {Mantelet},
  {Marchal}, {Marchant}, {Marconi}, {Marie}, {Marinoni}, {Marrese},
  {Marschalk{\'o}}, {Marshall}, {Mart{\'\i}n-Fleitas}, {Martino}, {Mary},
  {Matijevi{\v{c}}}, {Mazeh}, {McMillan}, {Messina}, {Mestre}, {Michalik},
  {Millar}, {Miranda}, {Molina}, {Molinaro}, {Molinaro}, {Moln{\'a}r},
  {Moniez}, {Montegriffo}, {Monteiro}, {Mor}, {Mora}, {Morbidelli}, {Morel},
  {Morgenthaler}, {Morley}, {Morris}, {Mulone}, {Muraveva}, {Musella},
  {Narbonne}, {Nelemans}, {Nicastro}, {Noval}, {Ord{\'e}novic},
  {Ordieres-Mer{\'e}}, {Osborne}, {Pagani}, {Pagano}, {Pailler}, {Palacin},
  {Palaversa}, {Parsons}, {Paulsen}, {Pecoraro}, {Pedrosa}, {Pentik{\"a}inen},
  {Pereira}, {Pichon}, {Piersimoni}, {Pineau}, {Plachy}, {Plum}, {Poujoulet},
  {Pr{\v{s}}a}, {Pulone}, {Ragaini}, {Rago}, {Rambaux}, {Ramos-Lerate},
  {Ranalli}, {Rauw}, {Read}, {Regibo}, {Renk}, {Reyl{\'e}}, {Ribeiro},
  {Rimoldini}, {Ripepi}, {Riva}, {Rixon}, {Roelens}, {Romero-G{\'o}mez},
  {Rowell}, {Royer}, {Rudolph}, {Ruiz-Dern}, {Sadowski}, {Sagrist{\`a}
  Sell{\'e}s}, {Sahlmann}, {Salgado}, {Salguero}, {Sarasso}, {Savietto},
  {Schnorhk}, {Schultheis}, {Sciacca}, {Segol}, {Segovia}, {Segransan},
  {Serpell}, {Shih}, {Smareglia}, {Smart}, {Smith}, {Solano}, {Solitro},
  {Sordo}, {Soria Nieto}, {Souchay}, {Spagna}, {Spoto}, {Stampa}, {Steele},
  {Steidelm{\"u}ller}, {Stephenson}, {Stoev}, {Suess}, {S{\"u}veges}, {Surdej},
  {Szabados}, {Szegedi-Elek}, {Tapiador}, {Taris}, {Tauran}, {Taylor},
  {Teixeira}, {Terrett}, {Tingley}, {Trager}, {Turon}, {Ulla}, {Utrilla},
  {Valentini}, {van Elteren}, {Van Hemelryck}, {van Leeuwen}, {Varadi},
  {Vecchiato}, {Veljanoski}, {Via}, {Vicente}, {Vogt}, {Voss}, {Votruba},
  {Voutsinas}, {Walmsley}, {Weiler}, {Weingrill}, {Werner}, {Wevers},
  {Whitehead}, {Wyrzykowski}, {Yoldas}, {{\v{Z}}erjal}, {Zucker}, {Zurbach},
  {Zwitter}, {Alecu}, {Allen}, {Allende Prieto}, {Amorim},
  {Anglada-Escud{\'e}}, {Arsenijevic}, {Azaz}, {Balm}, {Beck}, {Bernstein},
  {Bigot}, {Bijaoui}, {Blasco}, {Bonfigli}, {Bono}, {Boudreault}, {Bressan},
  {Brown}, {Brunet}, {Bunclark}, {Buonanno}, {Butkevich}, {Carret}, {Carrion},
  {Chemin}, {Ch{\'e}reau}, {Corcione}, {Darmigny}, {de Boer}, {de Teodoro}, {de
  Zeeuw}, {Delle Luche}, {Domingues}, {Dubath}, {Fodor}, {Fr{\'e}zouls},
  {Fries}, {Fustes}, {Fyfe}, {Gallardo}, {Gallegos}, {Gardiol}, {Gebran},
  {Gomboc}, {G{\'o}mez}, {Grux}, {Gueguen}, {Heyrovsky}, {Hoar}, {Iannicola},
  {Isasi Parache}, {Janotto}, {Joliet}, {Jonckheere}, {Keil}, {Kim},
  {Klagyivik}, {Klar}, {Knude}, {Kochukhov}, {Kolka}, {Kos}, {Kutka}, {Lainey},
  {LeBouquin}, {Liu}, {Loreggia}, {Makarov}, {Marseille}, {Martayan},
  {Martinez-Rubi}, {Massart}, {Meynadier}, {Mignot}, {Munari}, {Nguyen},
  {Nordlander}, {Ocvirk}, {O'Flaherty}, {Olias Sanz}, {Ortiz}, {Osorio},
  {Oszkiewicz}, {Ouzounis}, {Palmer}, {Park}, {Pasquato}, {Peltzer}, {Peralta},
  {P{\'e}turaud}, {Pieniluoma}, {Pigozzi}, {Poels}, {Prat}, {Prod'homme},
  {Raison}, {Rebordao}, {Risquez}, {Rocca-Volmerange}, {Rosen}, {Ruiz-Fuertes},
  {Russo}, {Sembay}, {Serraller Vizcaino}, {Short}, {Siebert}, {Silva},
  {Sinachopoulos}, {Slezak}, {Soffel}, {Sosnowska}, {Strai{\v{z}}ys}, {ter
  Linden}, {Terrell}, {Theil}, {Tiede}, {Troisi}, {Tsalmantza}, {Tur},
  {Vaccari}, {Vachier}, {Valles}, {Van Hamme}, {Veltz}, {Virtanen}, {Wallut},
  {Wichmann}, {Wilkinson}, {Ziaeepour}, \& {Zschocke}}]{gaiamission}
{Gaia Collaboration}, {Prusti}, T., {de Bruijne}, J.~H.~J., {et~al.} 2016,
  \aap, 595, A1

\bibitem[{{Gaze} \& {Shajn}(1951)}]{Gaze1951}
{Gaze}, V.~F. \& {Shajn}, G.~A. 1951, Izvestiya Ordena Trudovogo Krasnogo
  Znameni Krymskoj Astrofizicheskoj Observatorii, 7, 93

\bibitem[{{Gaze} \& {Shajn}(1955)}]{Gaze1955}
{Gaze}, V.~F. \& {Shajn}, G.~A. 1955, Izvestiya Ordena Trudovogo Krasnogo
  Znameni Krymskoj Astrofizicheskoj Observatorii, 15, 11

\bibitem[{{Green} {et~al.}(2019){Green}, {Schlafly}, {Zucker}, {Speagle}, \&
  {Finkbeiner}}]{2019arXiv190502734G}
{Green}, G.~M., {Schlafly}, E.~F., {Zucker}, C., {Speagle}, J.~S., \&
  {Finkbeiner}, D.~P. 2019, arXiv e-prints, arXiv:1905.02734

\bibitem[{{Henden} {et~al.}(2009){Henden}, {Welch}, {Terrell}, \&
  {Levine}}]{Henden2009}
{Henden}, A.~A., {Welch}, D.~L., {Terrell}, D., \& {Levine}, S.~E. 2009, in
  American Astronomical Society Meeting Abstracts, Vol. 214, American
  Astronomical Society Meeting Abstracts \#214, 407.02

\bibitem[{{Hindsley} \& {Harrington}(1994)}]{Hindsley1994AJ....107..280H}
{Hindsley}, R.~B. \& {Harrington}, R.~S. 1994, \aj, 107, 280

\bibitem[{{Israel} {et~al.}(2003){Israel}, {Kloppenburg}, {Dewdney}, \&
  {Bally}}]{israel2003}
{Israel}, F.~P., {Kloppenburg}, M., {Dewdney}, P.~E., \& {Bally}, J. 2003,
  \aap, 398, 1063

\bibitem[{{Jarrett} {et~al.}(2011){Jarrett}, {Cohen}, {Masci}, {Wright},
  {Stern}, {Benford}, {Blain}, {Carey}, {Cutri}, {Eisenhardt}, {Lonsdale},
  {Mainzer}, {Marsh}, {Padgett}, {Petty}, {Ressler}, {Skrutskie}, {Stanford},
  {Surace}, {Tsai}, {Wheelock}, \& {Yan}}]{Jarrett2011}
{Jarrett}, T.~H., {Cohen}, M., {Masci}, F., {et~al.} 2011, \apj, 735, 112

\bibitem[{Kennicutt {et~al.}(1995)Kennicutt, Bresolin, Bomans, Bothun, \&
  Thompson}]{Kennicutt:1994wu}
Kennicutt, R.~C., Bresolin, F., Bomans, D.~J., Bothun, G., \& Thompson, I.~B.
  1995, Astron. J., 109, 594

\bibitem[{{Kharchenko} \& {Roeser}(2009)}]{Kharchenko2009}
{Kharchenko}, N.~V. \& {Roeser}, S. 2009, VizieR Online Data Catalog, I/280B

\bibitem[{Lagrois {et~al.}(2012)Lagrois, Joncas, \& Drissen}]{Lagrois2012}
Lagrois, D., Joncas, G., \& Drissen, L. 2012, Monthly Notices of the Royal
  Astronomical Society, 420, 2280

\bibitem[{{Lasker} {et~al.}(2008){Lasker}, {Lattanzi}, {McLean}, {Bucciarelli},
  {Drimmel}, {Garcia}, {Greene}, {Guglielmetti}, {Hanley}, {Hawkins},
  {Laidler}, {Loomis}, {Meakes}, {Mignani}, {Morbidelli}, {Morrison},
  {Pannunzio}, {Rosenberg}, {Sarasso}, {Smart}, {Spagna}, {Sturch},
  {Volpicelli}, {White}, {Wolfe}, \& {Zacchei}}]{Lasker2008}
{Lasker}, B.~M., {Lattanzi}, M.~G., {McLean}, B.~J., {et~al.} 2008, \aj, 136,
  735

\bibitem[{Lee {et~al.}(1996)Lee, Snell, \& Dickman}]{Lee_1996}
Lee, Y., Snell, R.~L., \& Dickman, R.~L. 1996, The Astrophysical Journal, 472,
  275

\bibitem[{{Magnier} {et~al.}(2013){Magnier}, {Schlafly}, {Finkbeiner}, {Juric},
  {Tonry}, {Burgett}, {Chambers}, {Flewelling}, {Kaiser}, {Kudritzki},
  {Morgan}, {Price}, {Sweeney}, \& {Stubbs}}]{Magnier2013}
{Magnier}, E.~A., {Schlafly}, E., {Finkbeiner}, D., {et~al.} 2013, \apjs, 205,
  20

\bibitem[{{Mahy} {et~al.}(2015){Mahy}, {Rauw}, {De Becker}, {Eenens}, \&
  {Flores}}]{Mahy2015}
{Mahy}, L., {Rauw}, G., {De Becker}, M., {Eenens}, P., \& {Flores}, C.~A. 2015,
  \aap, 577, A23

\bibitem[{{McDonald} {et~al.}(2017){McDonald}, {Zijlstra}, \&
  {Watson}}]{McDonald2017MNRAS.471..770M}
{McDonald}, I., {Zijlstra}, A.~A., \& {Watson}, R.~A. 2017, \mnras, 471, 770

\bibitem[{Momcheva {et~al.}(2013)Momcheva, Lee, Ly, Salim, Dale, Ouchi, Finn,
  \& Ono}]{Momcheva_2013}
Momcheva, I.~G., Lee, J.~C., Ly, C., {et~al.} 2013, The Astronomical Journal,
  145, 47

\bibitem[{Neugebauer {et~al.}(1984)Neugebauer, Habing, Duinen, Aumann, Baud,
  Beichman, Beintema, Boggess, Clegg, De~Jong, Emerson, Gautier, Gillett,
  Harris, Hauser, Houck, Jennings, Low, Marsden, \& Young}]{Neugebauer_IRAS}
Neugebauer, G., Habing, H., Duinen, R., {et~al.} 1984, Astrophysical Journal,
  278, L1 - L6 (1984), 278

\bibitem[{{Ofek}(2008)}]{Ofek2008}
{Ofek}, E.~O. 2008, \pasp, 120, 1128

\bibitem[{{Osterbrock} \& {Ferland}(2006)}]{2006agna.book.....O}
{Osterbrock}, D.~E. \& {Ferland}, G.~J. 2006, {Astrophysics of gaseous nebulae
  and active galactic nuclei} (University Science Books)

\bibitem[{{Panagia}(1973)}]{1973AJ.....78..929P}
{Panagia}, N. 1973, \aj, 78, 929

\bibitem[{{Parker} {et~al.}(1979){Parker}, {Gull}, \& {Kirshner}}]{Parker1979}
{Parker}, R.~A.~R., {Gull}, T.~R., \& {Kirshner}, R.~P. 1979, {An emission-line
  survey of the Milky Way}, Vol. 434 (NASA-Washington : Scientific and
  Technical Inform)

\bibitem[{{Perryman} {et~al.}(1997){Perryman}, {Lindegren}, {Kovalevsky},
  {Hog}, {Bastian}, {Bernacca}, {Creze}, {Donati}, {Grenon}, {Grewing}, {van
  Leeuwen}, {van der Marel}, {Mignard}, {Murray}, {Le Poole}, {Schrijver},
  {Turon}, {Arenou}, {Froeschle}, \& {Petersen}}]{Perryman1997}
{Perryman}, M.~A.~C., {Lindegren}, L., {Kovalevsky}, J., {et~al.} 1997, \aap,
  500, 501

\bibitem[{{Pickles} \& {Depagne}(2010)}]{Pickles2010}
{Pickles}, A. \& {Depagne}, {\'E}. 2010, \pasp, 122, 1437

\bibitem[{{Pipenbrink} \& {Wendker}(1988)}]{Pipenbrink1988}
{Pipenbrink}, A. \& {Wendker}, H.~J. 1988, \aap, 191, 313

\bibitem[{{Poole} {et~al.}(2008){Poole}, {Breeveld}, {Page}, {Land sman},
  {Holland}, {Roming}, {Kuin}, {Brown}, {Gronwall}, {Hunsberger}, {Koch},
  {Mason}, {Schady}, {vanden Berk}, {Blustin}, {Boyd}, {Broos}, {Carter},
  {Chester}, {Cucchiara}, {Hancock}, {Huckle}, {Immler}, {Ivanushkina},
  {Kennedy}, {Marshall}, {Morgan}, {Pandey}, {de Pasquale}, {Smith}, \&
  {Still}}]{2008MNRAS.383..627P}
{Poole}, T.~S., {Breeveld}, A.~A., {Page}, M.~J., {et~al.} 2008, \mnras, 383,
  627

\bibitem[{{Pourbaix} {et~al.}(2004){Pourbaix}, {Tokovinin, A. A.}, {Batten, A.
  H.}, {Fekel, F. C.}, {Hartkopf, W. I.}, {Levato, H.}, {Morrell, N. I.},
  {Torres, G.}, \& {Udry, S.}}]{Pourbaix2004}
{Pourbaix}, D., {Tokovinin, A. A.}, {Batten, A. H.}, {et~al.} 2004, A\&A, 424,
  727

\bibitem[{{Price} {et~al.}(2010){Price}, {Smith}, {Kuchar}, {Mizuno}, \&
  {Kraemer}}]{Price2010ApJS..190..203P}
{Price}, S.~D., {Smith}, B.~J., {Kuchar}, T.~A., {Mizuno}, D.~R., \& {Kraemer},
  K.~E. 2010, \apjs, 190, 203

\bibitem[{{Rate} \& {Crowther}(2020)}]{rate2020}
{Rate}, G. \& {Crowther}, P.~A. 2020, \mnras, 493, 1512

\bibitem[{{Rygl} {et~al.}(2012){Rygl}, {Brunthaler}, {Sanna}, {Menten}, {Reid},
  {van Langevelde}, {Honma}, {Torstensson}, \& {Fujisawa}}]{Rygl2012}
{Rygl}, K.~L.~J., {Brunthaler}, A., {Sanna}, A., {et~al.} 2012, \aap, 539, A79

\bibitem[{{Schirmer}(2013)}]{SchirmerTheli}
{Schirmer}, M. 2013, \apjs, 209, 21

\bibitem[{{Schlafly} \& {Finkbeiner}(2011)}]{Schlafly2011}
{Schlafly}, E.~F. \& {Finkbeiner}, D.~P. 2011, \apj, 737, 103

\bibitem[{Skrutskie {et~al.}(2006)Skrutskie, Cutri, Stiening, Weinberg,
  Schneider, Carpenter, Beichman, Capps, Chester, Elias,
  {et~al.}}]{skrutskie2006two}
Skrutskie, M., Cutri, R., Stiening, R., {et~al.} 2006, The Astronomical
  Journal, 131, 1163

\bibitem[{{Smith} {et~al.}(2012){Smith}, {Dougherty}, \& {Beasley}}]{smith2012}
{Smith}, I.~A., {Dougherty}, S.~M., \& {Beasley}, A.~J. 2012, in Astronomical
  Society of the Pacific Conference Series, Vol. 465, Proceedings of a
  Scientific Meeting in Honor of Anthony F. J. Moffat, ed. L.~{Drissen},
  C.~{Robert}, N.~{St-Louis}, \& A.~F.~J. {Moffat}, 336

\bibitem[{Sternberg {et~al.}(2003)Sternberg, Hoffmann, \&
  Pauldrach}]{Sternberg_2003}
Sternberg, A., Hoffmann, T.~L., \& Pauldrach, A. W.~A. 2003, The Astrophysical
  Journal, 599, 1333–1343

\bibitem[{{Takita} {et~al.}(2010){Takita}, {Kataza}, {Kitamura}, {Ishihara},
  {Ita}, {Oyabu}, \& {Ueno}}]{Takita2010}
{Takita}, S., {Kataza}, H., {Kitamura}, Y., {et~al.} 2010, \aap, 519, A83

\bibitem[{{Uyan{\i}ker} {et~al.}(2001){Uyan{\i}ker}, {F{\"u}rst}, {Reich},
  {Aschenbach}, \& {Wielebinski}}]{uyaniker2001}
{Uyan{\i}ker}, B., {F{\"u}rst}, E., {Reich}, W., {Aschenbach}, B., \&
  {Wielebinski}, R. 2001, \aap, 371, 675

\bibitem[{{van der Tak} {et~al.}(2007){van der Tak}, {Black}, {Sch{\"o}ier},
  {Jansen}, \& {van Dishoeck}}]{vanderTak2007}
{van der Tak}, F.~F.~S., {Black}, J.~H., {Sch{\"o}ier}, F.~L., {Jansen}, D.~J.,
  \& {van Dishoeck}, E.~F. 2007, \aap, 468, 627

\bibitem[{{van Dokkum} {et~al.}(2012){van Dokkum}, {Bloom}, \&
  {Tewes}}]{2012ascl.soft07005V}
{van Dokkum}, P.~G., {Bloom}, J., \& {Tewes}, M. 2012, {L.A.Cosmic: Laplacian
  Cosmic Ray Identification}, Astrophysics Source Code Library

\bibitem[{{Vioque, M.} {et~al.}(2020){Vioque, M.}, {Oudmaijer, R. D.},
  {Schreiner, M.}, {Mendigut\'{\i}a, I.}, {Baines, D.}, {Mowlavi, N.}, \&
  {P\'erez-Mart\'{\i}nez, R.}}]{Vioque2020}
{Vioque, M.}, {Oudmaijer, R. D.}, {Schreiner, M.}, {et~al.} 2020, A\&A, 638,
  A21

\bibitem[{{Wenger} {et~al.}(2000){Wenger}, {Ochsenbein}, {Egret}, {Dubois},
  {Bonnarel}, {Borde}, {Genova}, {Jasniewicz}, {Lalo{\"e}}, {Lesteven}, \&
  {Monier}}]{simbad2000}
{Wenger}, M., {Ochsenbein}, F., {Egret}, D., {et~al.} 2000, \aaps, 143, 9

\bibitem[{Whittet(2002)}]{whittet2002dust}
Whittet, D.~C. 2002, Dust in the galactic environment (CRC press)

\bibitem[{{Williams}(2011)}]{Williams2011}
{Williams}, P. 2011, Bulletin de la Societe Royale des Sciences de Liege, 80,
  595

\bibitem[{Wood \& Churchwell(1989)}]{wood1989massive}
Wood, D.~O. \& Churchwell, E. 1989, The Astrophysical Journal, 340, 265

\bibitem[{{Wright} {et~al.}(2010){Wright}, {Eisenhardt}, {Mainzer}, {Ressler},
  {Cutri}, {Jarrett}, {Kirkpatrick}, {Padgett}, {McMillan}, {Skrutskie},
  {Stanford}, {Cohen}, {Walker}, {Mather}, {Leisawitz}, {Gautier}, {McLean},
  {Benford}, {Lonsdale}, {Blain}, {Mendez}, {Irace}, {Duval}, {Liu}, {Royer},
  {Heinrichsen}, {Howard}, {Shannon}, {Kendall}, {Walsh}, {Larsen}, {Cardon},
  {Schick}, {Schwalm}, {Abid}, {Fabinsky}, {Naes}, \& {Tsai}}]{Wright2010}
{Wright}, E.~L., {Eisenhardt}, P. R.~M., {Mainzer}, A.~K., {et~al.} 2010, \aj,
  140, 1868

\bibitem[{Zhu {et~al.}(2017)Zhu, Tian, Li, \& Zhang}]{zhu2017_gastoextinction}
Zhu, H., Tian, W., Li, A., \& Zhang, M. 2017, Monthly Notices of the Royal
  Astronomical Society, 471, 3494

\end{thebibliography}

\appendix
\section{Stellar neighborhood}
\par
\begin{figure}
\includegraphics[width=0.48\textwidth]{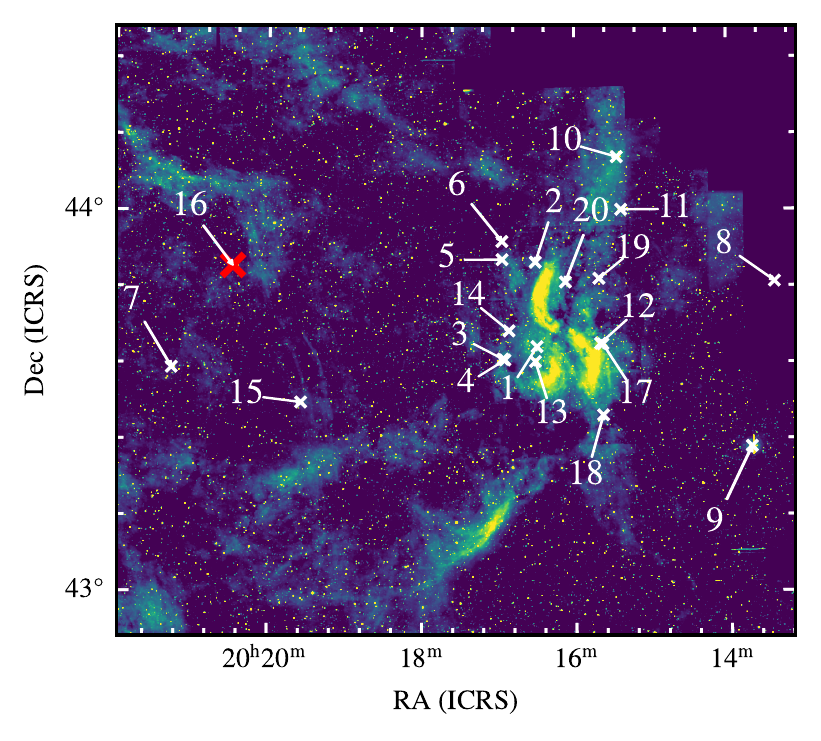}
\centering
\caption{Identification chart of bright stars near Simeis~57 for which
  continuum spectra are shown in Fig\,\ref{fig:stellar_spectra}. The
  red cross denotes star No. 16 (HD 193793/WR140 aka the spectroscopic
  binary SBC9 1232 \citep{Pourbaix2004}), which is the most likely
  candidate for the exciting star of Simeis~57. Background map is as
  in Fig.  \ref{fig:Ha_mosaic}.}
\label{fig:bright_stars}
\end{figure}

\begin{figure}
\includegraphics[width=0.98\textwidth]{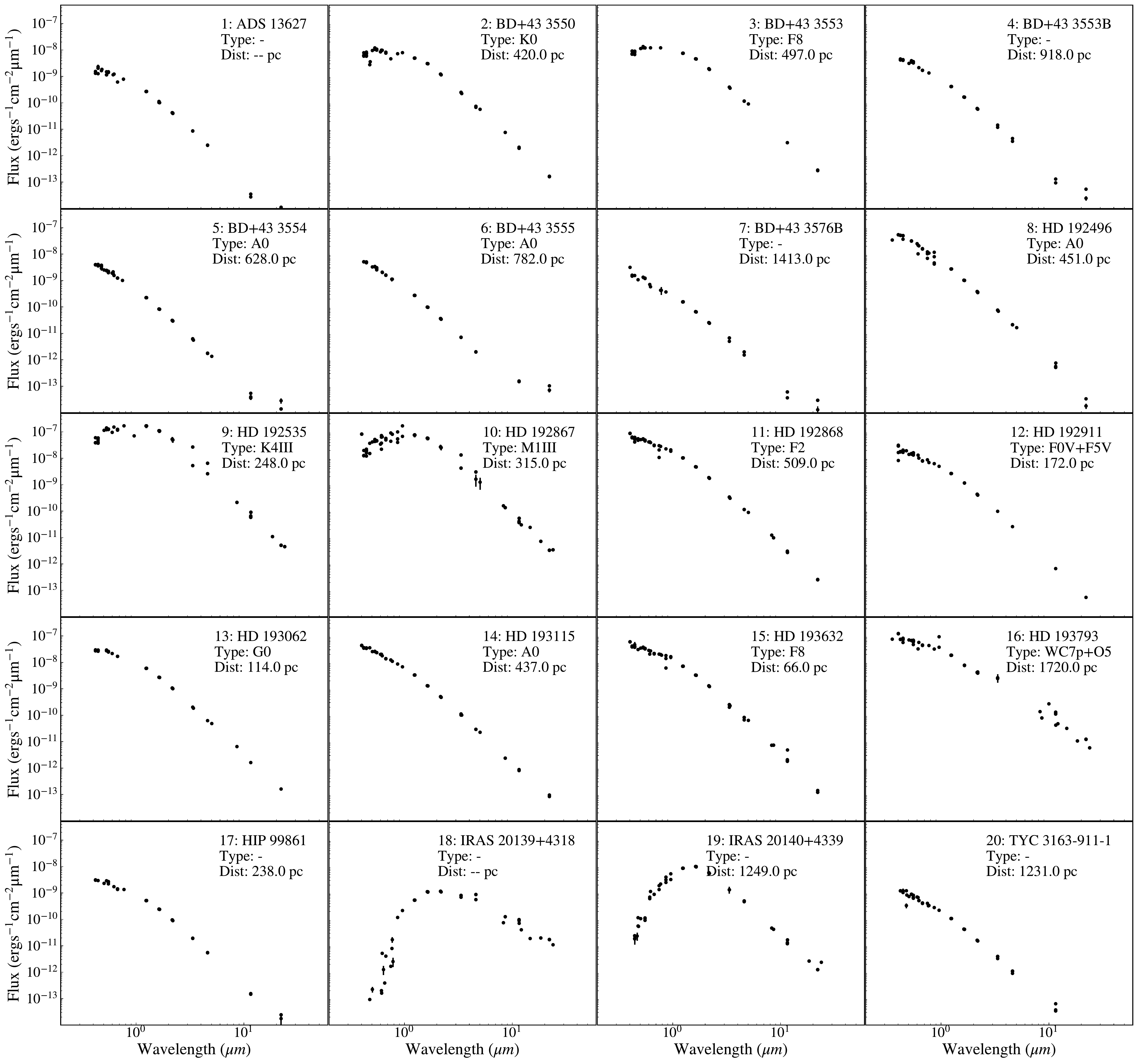}
\centering
\caption{Continuum spectra of stars identified in
  Fig\,\ref{fig:bright_stars}. The distanced are as available in 
  Simbad \citep{simbad2000}}.
\label{fig:stellar_spectra}
\end{figure}

 The continuum spectra of 20 relatively prominent stars near
Simeis~57, identified in Fig\,\ref{fig:bright_stars}, are shown in
Fig.\,\ref{fig:stellar_spectra}. Thirteen of these stars are also
contained in the UV-to-FIR images from Fig.\,\ref{fig:image_spectrum}.
The stellar fluxes were taken from Simbad \citep{simbad2000}, which
was accessed through the CDS portal.  Photometric measurements are
mostly part of all-sky surveys. Catalogs and instruments included
in the data are as follows: the Sloan Digital Sky Survey (SDSS)
\citep{Aguado2019}, the 2MASS All-Sky Catalog of Point Sources
\citep{Cutri2003}, WISE \citep{Wright2010,Jarrett2011}, Tycho-2 and
the Tycho Input Catalogue \cite{Pickles2010,Egret1992A,Ofek2008}, the
MSX6C Infrared Point Source Catalog \citep{Egan2003}, AKARI
\citep{Takita2010}, Pan-STARRS \citep{Magnier2013}, GAIA
\citep{gaiamission,gaia_data}, the All-Sky Compiled Catalogue
\citep{Kharchenko2009}, Hipparcos \citep{Perryman1997}, The AAVSO
Photometric All-Sky Survey (APASS) \citep{Henden2009}, the IRAS
PSC/FSC Combined Catalogue \citep{Abrahamyan2015A}, the IPHAS DR2
Source Catalogue \citep{Barentsen2014}, the (Second-Generation) Guide
Star Catalog \citep{Lasker2008}, the Catalogue of new Herbig Ae/Be and
classical Be stars \citep{Vioque2020}, DIRBE Near-infrared Stellar
Light Curves \citep{Price2010ApJS..190..203P}, the U.S. Naval
Observatory Catalog of Positions of Infrared Stellar Sources
\citep{Hindsley1994AJ....107..280H}, and Parameters and IR excesses of
Gaia DR1 stars \citep{McDonald2017MNRAS.471..770M}.
\par For many stars, a small fraction of the photometric fluxes
returned by CDS portal fall significantly below the stellar
spectrum. In almost all cases, these photometric data points are part
of the Pan-STARRS catalog. For the same wavelength, multiple (up to
four) photometric data points are returned with different fluxes. For
each of the Pan-STARRS filters, one of the photometric points is
consistent with the stellar spectrum as defined by data from other
catalogs and GAIA. In these cases, we deleted the other points.
For two stars, the photometric fluxes in the IPHAS Source Catalogue
were significantly below those other compilations at similar
wavelengths, and in particular below the Gaia measurements. Since the
accuracy of Gaia is higher than that of the IPHAS catalog, we also 
removed the IPHAS data points for these two stars.

Four of the stars (Nos. 7, 16, 19, and 20) are at distances close to
the estimated distance of Simeis~57, whereas the distance of two stars
(Nos. 1 and 18) is unknown.  Stars Nos. 2, 3, 9, 10, 11, 12, 13, 15,
and 17 are all too nearby and have spectral types too late to be of
interest here.  Stars Nos. 5, 6, 8, and 14 are relatively luminous
stars of early type A0, but their output of ionizing UV photons is
still too small to matter nor are they sufficiently distant. The
spectral type of stars Nos. 1, 4, 7, 18, 19, and 20 is unknown. The
IRAS PSC stars Nos. 18 and 19, prominent in the WISE images of
Fig.\,\ref{fig:image_spectrum}, appear to be dust-embedded stars of
relatively low luminosity. This leaves for consideration stars Nos. 1,
4, 7, 16, and 20. Only star No. 16 (HD 193793) seems bright enough.
As the spectroscopic binary SBC9 1232 \citep{Pourbaix2004}, it
consists of an evolved O4/O5 supergiant and the WC7p Wolf-Rayet star WR~140
at a distance of 1.7 kpc with a combined luminosity of about
$2\times10^6$ L$_{\odot}$.  \onecolumn

\end{document}